\documentclass{aa}

\usepackage{placeins}
\usepackage{graphicx}
\usepackage{txfonts}
\usepackage{xcolor}
\newcommand{\micron}{$\rm \mu$m}
\newcommand{\massyear}{M$_\odot$~yr$^{-1}$}

\usepackage{amsmath}
\usepackage{bm}
\usepackage{subfig}
\usepackage{hyperref}
\usepackage[normalem]{ulem}

\begin{document} 

   \title{An interferometric mid-infrared study of the eruptive star binary Z~CMa with MATISSE/VLTI\thanks{Reproduced with permission from Astronomy \& Astrophysics, \textcopyright ~ESO.}}

   \subtitle{I. Imaging the protoplanetary disks during the 2023 outburst\thanks{Based on observations collected at the European Organisation for Astronomical Research in the Southern Hemisphere under ESO programs 0106.C-0501(B), 0108.C-0385(D), and 0110.C-4209(A).}}

 \author{F.~Lykou \inst{1,2}
          \and
          J.~Varga\inst{1,2}
          \and
          F.~Cruz-Sa\'enz de Miera\inst{3,1,2}
        \and
          P.~\'Abrah\'am\inst{1,2,4,5}
        \and
        \'A.~K\'osp\'al\inst{1,2,4,6}
        \and
        B.~Lopez\inst{7}
        \and
        T.~Henning\inst{6}
        \and
        S.~Wolf\inst{8}
        \and
        G.~Weigelt\inst{9}
        \and
        F.~Millour\inst{7}
        \and
        M.~Hogerheijde\inst{10}
        \and
        L.~Chen \inst{1,2}
        \and
        T.~Ratzka \inst{11}
        \and
        W.~Danchi \inst{12}
        \and
        P.~Boley \inst{6}
        \and
        J.-C.~Augereau\inst{13}
        \and
        P.~Priolet \inst{13}
                }

   \institute{Konkoly Observatory, HUN-REN Research Centre for Astronomy and Earth Sciences, Konkoly Thege Mikl\'os \'ut 15-17, 1121 Budapest, Hungary 
        \and
       MTA Centre of Excellence, HUN-REN CSFK, Budapest, Konkoly Thege Mikl\'os \'ut 15-17, 1121 Budapest, Hungary
         \and
         Institut de Recherche en Astrophysique et Planétologie, Université de Toulouse, UT3-PS, OMP, CNRS, 9 av. du Colonel Roche, 31028 Toulouse Cedex 4, France
        \and
        Institute of Physics and Astronomy, ELTE E\"otv\"os Lor\'and University, P\'azm\'any P\'eter s\'et\'any 1/A, 1117 Budapest, Hungary
         \and
         Department of Astrophysics, University of Vienna, T\"urkenschanzstrasse 17, 1180 Vienna, Austria
         \and
          Max Planck Institute for Astronomy, K\"onigstuhl 17, D-69117 Heidelberg, Germany
         \and
         Laboratoire Lagrange, Université C\^ote d’Azur, Observatoire de la C\^ote d’Azur, CNRS, Boulevard de l’Observatoire, CS 34229, 06304, Nice Cedex 4, France
         \and
         Institut f\"ur Theoretische Physik und Astrophysik, Christian-Albrechts-Universit\"at zu Kiel, Leibnizstraße 15, 24118, Kiel, Germany
         \and
         Max-Planck-Institut f\"ur Radioastronomie, Auf dem H\"ugel 69, 53121 Bonn, Germany
                \and
         Leiden Observatory, Leiden University, P.O. Box 9513, NL2300, RA Leiden, The Netherlands
         \and
         Planetarium im Museum am Sch\"olerberg, Klaus-Strick-Weg 10, 49082, Osnabr\"uck, Germany 
         \and
      NASA Goddard Space Flight Center, Astrophysics Division, Greenbelt, MD 20771, USA
        \and
         Univ. Grenoble Alpes, CNRS, IPAG, 38000 Grenoble, France
         }
   \date{Received 10 December 2024; accepted 13 August, 2025}

 
  \abstract
   {}
   {The current work is part of a series aimed at producing the first ever mid-infrared images of protoplanetary disks in the binary system of eruptive stars Z~CMa and studying their individual properties.}
   {We obtained high-angular-resolution interferometric observations with MATISSE/VLTI in the $L$ (2.9 -- 4.1 \micron), $M$ (4.5 -- 4.9 \micron), and $N$ (8 -- 13 \micron) bands, as well as spectroscopic observations in the near-infrared (NIR) with SpeX/IRTF. 
   We present our quantitative analysis on the interferometric data using geometric model fitting, image reconstruction algorithms, and orbital simulation tools, and we compare our findings to those of literature studies.}
   {The mid-infrared (MIR) emitting regions of the individual protoplanetary disks in the binary system Z~CMa are resolved by MATISSE/VLTI. The observations were obtained during a serendipitous large outburst of the Herbig (HBe) star that lasted more than 100~days, while the FU~Orionis-type (FUor) companion is presumed to be in quiescence. The size of the MIR-emitting disk region of the more massive HBe star increases toward longer wavelengths from $<14$~mas at 3.5 \micron\ to $\ll 50$~mas at 11.5 \micron . The lack of substructures in the HBe disk might suggest that it is a continuous disk; however, this could be due to observational constraints. We also note a radial variation of the silicate absorption feature over the disk, where the optical depth increases inwards of $<$40~au radii. This contradicts the scenario of a carved, dusty cocoon surrounding the HBe star. In the case of the less massive FUor companion, the MIR-emitting region is much smaller with an angular size $\leq$15~mas (or else a physical radius $<9$ au) in all bands, suggesting a compact disk. Both disks are aligned within uncertainties, and their orientation agrees with that of the known jets. Furthermore, MATISSE data place the binary's separation at $117.88 \pm 0.73$~mas and a position angle of 139.16\degr\ $\pm$ 0.29\degr\ east of north. Our estimates for the orbital elements gave an eccentric orbit ($e\sim0.17$) with a moderate inclination ($i\sim 66$\degr). The derived total mass is $M_{\rm total} = 16.4^{+2.1}_{-2.3}$~M$_\odot$, while the period is approximately 950 years.}
   {Our MATISSE imaging of the Herbig disk during outburst indicates a temperature gradient for the disk, while imaging of the FUor companion's disk corroborates previous studies showing that FUor disks are rather compact in the MIR. We cannot infer any misalignment between the MATISSE results and earlier ALMA/JVLA data, nor can we infer any influence from the alleged flyby event.
   }


\titlerunning{Z~CMa with MATISSE/VLTI. I} 
   \maketitle
\nolinenumbers
\section{Introduction}

Dust and gas inside protoplanetary disks are the basic material, which will be reprocessed at later evolutionary stages, to form terrestrial and gaseous planets. One of the intermediary stages in star formation is the Class II young stellar objects (YSOs) \citep[e.g.,][]{lada1987,williams2011}. These are variable in the optical and near-infrared (NIR) due to eruptive events (outbursts) followed by changes in the rates at which they accrete mass from their protoplanetary disks. Typical mass-accretion rates range from $10^{-8}$ to $10^{-4}$~\massyear. These stars can be classified in different groups based on the strength and duration of said variability (see, e.g., \citealt{fischer} for a recent review on the subject matter). Prominent sub-groups are the FUor and the EXor YSOs, so named after their archetypes, FU Ori and EX Lup, respectively. Stars from these sub-groups are typically on the lower-mass end ($\leq 2$~M$_\odot$). Their eruptions can be in the range of 1 to 6 mag in the optical, and such outbursts last for several days to years (EXors), or even decades (FUors). On the other hand, the intermediate-mass YSOs (2 -- 12~M$_\odot$) also show variability in the form of (out)bursts, although these are typically of lower amplitude; that is, $\leq 2$~mag in the optical. The class of Herbig Ae/Be eruptive stars belongs in this category \citep{herbig1960}. In all of the above-mentioned cases, the eruptions are irregular.

There are currently two predominant theories as to what causes eruptions. On the one hand, they may result from physical changes -- instabilities and/or fragmentations -- within the protoplanetary disk itself \citep{2010ApJ...714L.133V,2024A&A...683A.202V,naya2024}. On the other hand, the orbital motion of a companion or a flyby may perturb the disk \citep{bonnell1992,2023ARep...67.1401S}. Regardless of the cause, the eruptions occur at the innermost regions of the protoplanetary disks within a distance $\leq1$~au from the protostar. In particular, these protoplanetary disks are thought to have a defined structure: (1) a hot and/or warm inner part emitting in the infrared within $\leq10$~au of the protostar, and (2) a cooler exterior emitting at far-IR and submillimeter wavelengths \citep[e.g.,][and references therein]{dullemond2010}. Their composition is a mixture of gas, dust, and ices. During an eruption, infalling gas within a few stellar radii becomes optically thick and is heated, which makes it visible in the optical and NIR. This is a major difference between FUor disks and those of quiescent protostars, since the latter have an optically thin gaseous disk and their circumstellar emission is dominated by their dusty component. Herbig Ae/Be star protoplanetary disks are overall more massive and much larger in size compared to typical FUor and T~Tauri disks. They are divided in two main groups depending on whether they are flared (Group I) or flat (Group II), while they can also have several other characteristics such as line emission, disk gaps, NIR and/or mid-infrared (MIR) excess, and processed dust material such as crystalline silicates or PAHs \citep{meeus2001,maaskant2013,menu2015}. Such small spatial scales of a few astronomical units can only be resolved by high-angular-resolution instruments, such as the Atacama Large Millimeter/submillimeter Array (ALMA) for radii of $\ge$10~au, and the Very Large Telescope Interferometer (VLTI) for shorter radii closer to the protostars. Multiwavelength observations are most suited to painting a clear picture of the processes in these disks.

\begin{figure*}[bthp]
    \centering
    \includegraphics[width=0.9\linewidth]{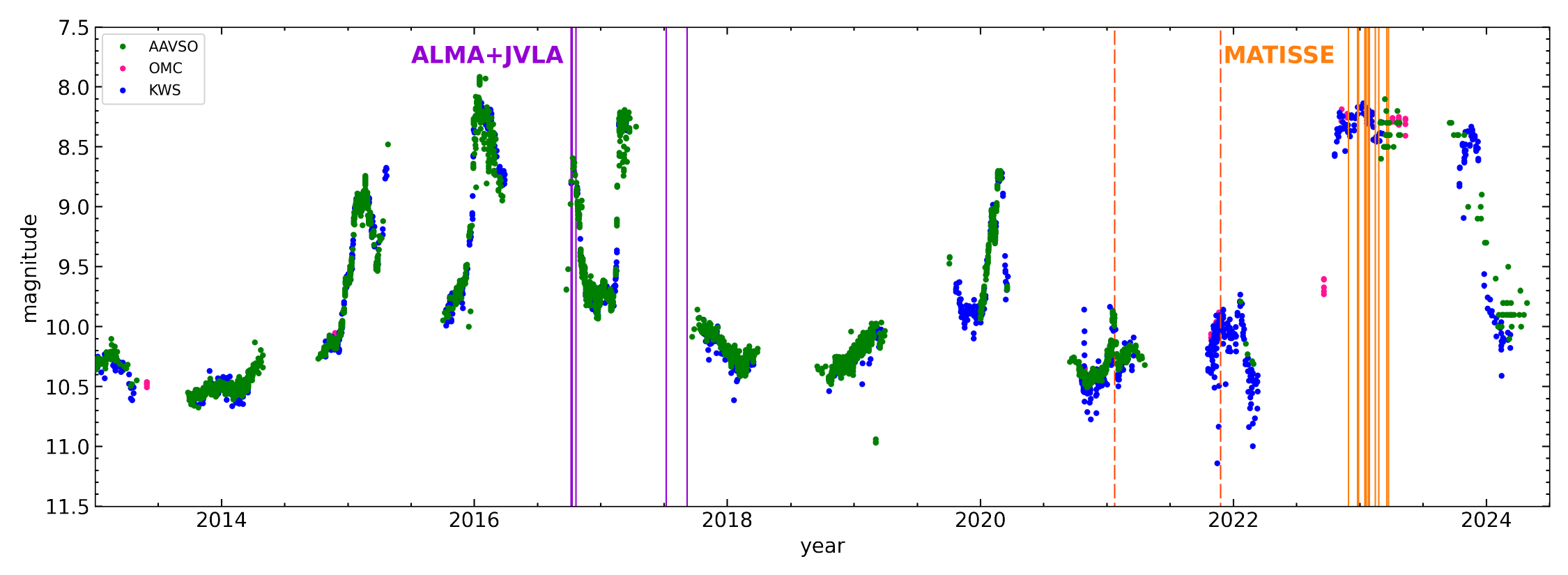}
     \caption{$V$-band light curve of Z~CMa with corresponding epochs of MATISSE observations (orange lines; dashed and solid for the quiescent and the outburst phase, respectively) as in Table~\ref{tab:matlog}. Photometric data from the Archive of the American Association of Variable Star Observers (AAVSO; green circles), the INTErnational Gamma-Ray Astrophysics Laboratory (INTEGRAL) Optical Monitoring Camera (OMC; magenta circles), and the Kamogata/Kiso/Kyoto Wide-field Survey of Variable Star Observers League in Japan \citep[KWS; blue circles][]{vsolj}. The imaging data were obtained during the outburst from late 2022 to early 2023. Also shown for reference are the epochs of archival ALMA and JVLA observations \citep[purple lines;][]{takami2019}.}
    \label{fig:vislc}
\end{figure*}

Z~CMa is a member of the Canis Majoris R1 star forming region \citep[e.g.,][]{1999MNRAS.310..210S} with a probable age of the order of 5~Myr \citep{2009A&A...506..711G}. Hereafter, we adopt the distance estimate of \citet{dong2022} ($1125\pm30$ pc), which derived a median distance based on Gaia data of nearby CMa R1 cluster members. Z~CMa is a binary with a separation of $\sim$0.1\arcsec\ at a position angle of PA$\sim$130\degr \citep{koresko1991, barth1994}. The northwest component (Z~CMa NW) is a Herbig star (HBe star), while the southeastern companion (Z~CMa SE) is an FUor. Constraining their individual masses has been a matter of debate. \citet{vdancker2004} estimated masses at $\sim 16$~M$_\odot$ for the HBe star and $\sim 3$~M$_\odot$ for the FUor. Both values are slightly higher than the assumed range for low- and intermediate-mass stars. On the other hand, \citet{fairlamb2015} placed the mass of the HBe slightly lower at $11\pm1.7$~M$_\odot$. The system is highly variable in the optical with several short-term ($\sim$100 days) bursts and long-term ($\sim$1000 days) outbursts that raise its brightness by $\delta V \sim 3$ mag (Fig.~\ref{fig:vislc}). This EXor-like variability is attributed to the Herbig star, while it is possible that the FUor companion has returned to quiescence (\citealt{sag2020} and references therein).

Near-infrared scattered-light imaging at large spatial scales (i.e., $>1$\arcsec ) revealed a streamer extending south-southwest from the binary system \citep{canovas2012, canovas2015, liu2016}. \citet{dong2022} argued that a potential flyby intruder is located at the edge of this streamer; however, \citet{zurlo} showed that the streamer is in fact part of a large-scale outflow. Further adding to the system's complexity, jets emanate from both stars. \citet{whelan2010} estimated the jet position angles at $\sim$235\degr\ and $\sim$245\degr\ for the FUor and HBe jet, respectively, although the uncertainty in the measurements is somewhat large due to the clumpiness seen at sub-arcsecond scales. This clumpiness is evident in both jets, and it could be explained by jet wiggles potentially instigated by additional companions in narrow orbits around the known stars \citep{whelan2010, antoniucci2016}. On the other hand, a disruption induced by a stellar flyby could also explain the jet wiggles.

Near-infrared interferometric observations revealed two very compact sources with a diameter of $\le 5$~au, and these correspond to the hot, innermost disk regions \citep{monnier2005,millangabet2006,benisty2010}. Observations of the system in the MIR have been limited by the spatial resolution of single-dish telescopes since the binary is seen as an oblate source \citep{malbet1993,2007AJ....133.1690P}, although the Herbig disk is the brightest of the two \citep{monnier2009}. On the other hand, the binary was resolved in the MIR with the MID-infrared Interferometric instrument (MIDI) on the VLTI, but the interpretation of the results was model-dependent \citep{ratzkaphd, varga2018}.

Protoplanetary disks often exhibit substructures such as spiral arms, arcs, rings, and gaps, to name but a few. It is widely believed that such structures are (in)direct evidence of disk-planet interactions \citep{benistyPP7}. For the case of Herbig disks, it is believed that the formation of gaps might be related to their evolutionary stage \citep{maaskant2013,menu2015}. The substructures have been observed at different spatial scales and wavelengths, and comparisons have shown that in several cases, the cooler outer disks observed in the (sub)millimeter continuum can be misaligned with structures observed in the NIR \citep[e.g.,][]{andrews2018,bohn2022}. Such misalignments might be more profound in Herbig disks \citep[e.g.,][]{gwori}, but this could be an observational bias since these disks are usually larger in size.

We revisited Z~CMa in the MIR with the Multi-AperTure mid-Infrared SpectroScopic Experiment \citep[MATISSE; ][]{matisse}, the new imaging instrument of the VLTI. Our goal was to image both disks and look for substructures within each, as well as for potential misalignments with respect to structures observed in the (sub)millimeter and NIR regimes. Furthermore, we explored the effect EXor-like (out)bursts can have on the inner disk region of the Herbig star. We also measured the sizes of the dusty, inner regions of each disk and, when possible, explored their mineralogical composition. This work focuses on the imaging results and is the first in a series of papers on the MATISSE data, which will be followed by a spatio-kinematic analysis of the gaseous inner regions and a study on the system's variability.

The observations are described in Section~\ref{sec:obs}, and the results of our analysis are shown in Section~\ref{sec:res}. Our interpretation is discussed in the penultimate section (Sect. \ref{sec:dis}), which is followed by our concluding remarks. Supplementary material can be found in the appendices.

\begin{figure}
    \centering
    \includegraphics[width=\columnwidth]{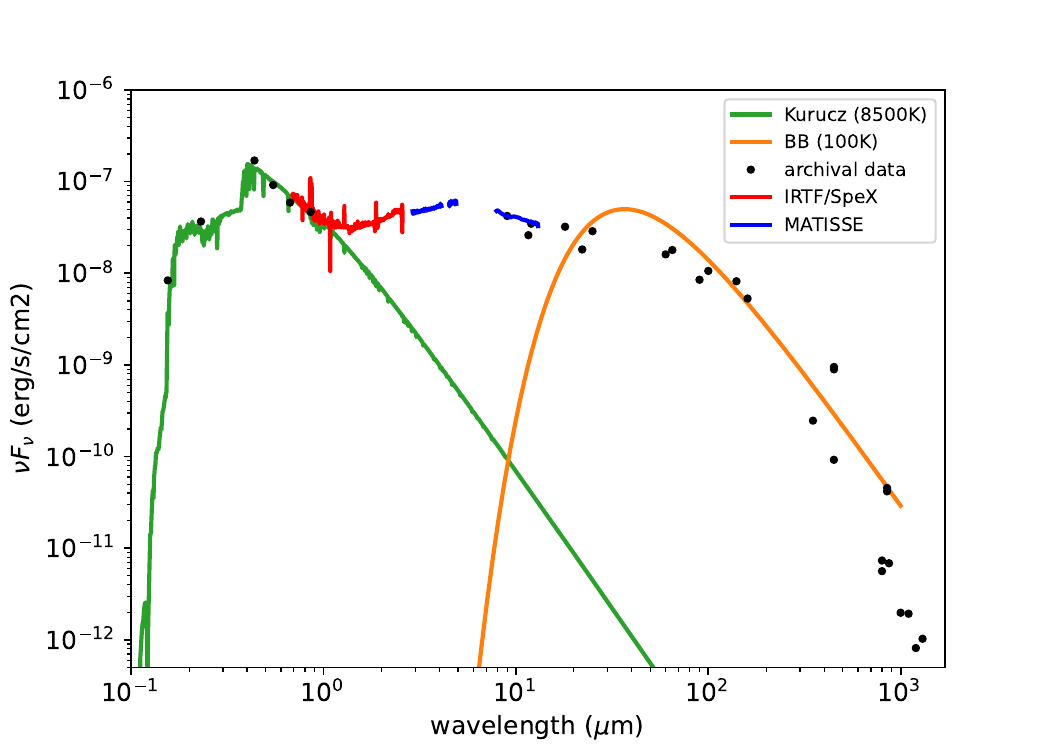}
    \caption{Dereddened SED of Z~CMa during the 2023 outburst. Our own SpeX/IRTF and MATISSE spectra are shown in red and blue, respectively. For a description of the archival photometry (black points), we refer the reader to Sect.~\ref{phot}. A Kurucz model (green) and a blackbody model (orange) are shown for reference. }
    \label{fig:specsed}
\end{figure}

\section{Observations} \label{sec:obs}

\subsection{Archival photometry and the variability of Z~CMa}\label{phot}

A current visual and $V$-band light curve of Z~CMa is shown in Fig.~\ref{fig:vislc}. The Herbig star experienced yet another outburst in late 2022 that lasted longer than 100 days. It appears to have remained in outburst until early December 2023, with a probable short-term dimming in October, although we note that there is a broad gap in temporal coverage from May to September 2023 when Z~CMa is not observable. As described below, this last outburst serendipitously occurred while our observations were obtained (orange lines). Earlier submillimeter and radio interferometric observations by ALMA and JVLA \citep{takami2019} are also indicated for reference (purple lines; Fig.~\ref{fig:vislc}). 

Figure~\ref{fig:specsed} shows the spectral energy distribution (SED) of Z~CMa during the outburst. The SED has been dereddened for $A_V=2.5$ (Appendix~\ref{ism}), and it encompasses emission from both sources in Z~CMa; however, the Herbig star and its disk would be the main contributors during the outburst. The optical photometry ($BVRI$; black points) represents the median values from the Kamogata/Kiso/Kyoto Wide-field Survey of Variable Star Observers League in Japan \citep[KWS,][]{vsolj} during the outburst. Our SpeX/IRTF and MATISSE spectra (see below) are shown in red and blue, respectively. Also included in black are archival data from Galaxy Evolution Explorer \citep[GALEX,][]{galexsynth}, Midsource Space Experiment \citep[MSX,][]{msx}, AKARI \citep{akariIRC}, Infrared Astronomical Satellite \citep[IRAS,][]{iras}, James Clerk Maxwell Telescope \citep[JCMT,][]{1991ApJ...382..270W, 1998MNRAS.301.1049D, 2008ApJS..175..277D}, Swedish-ESO Submillimetre Telescope \citep[SEST,][]{1993A&A...273..221R}, and Institute for Radio Astronomy in Millimetre Range 30-m telescope \citep[IRAM30,][]{1994A&A...281..161A}. A Kurucz model (8500~K; green) and a blackbody (100~K; orange) have been fit for reference. The far-infrared to (sub)millimeter tail of the system's SED has a spectral index of $\alpha = 3.2\pm0.2$.

\begin{figure}
    \centering
    \includegraphics[width=0.8\columnwidth]{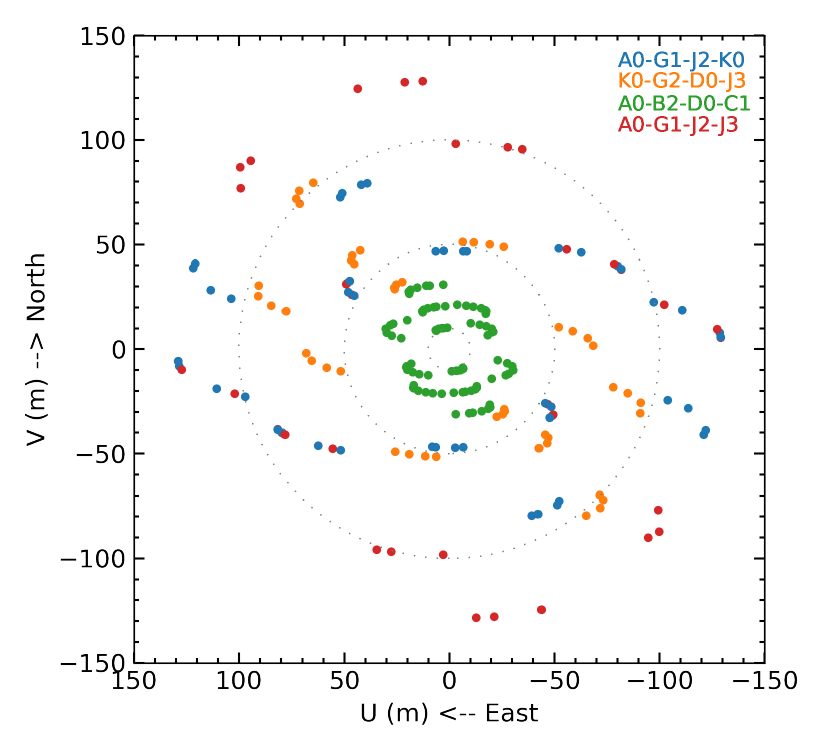}
    \caption{MATISSE $uv$-plane coverage of "imaging" mode observations of Z~CMa during its outburst phase. The $uv$ coordinates are colored per respective VLTI configuration array as in Table~\ref{tab:matlog}; that is, the small array included the A0-B2-D0-C1 quadruple, the medium array included the K0-G2-D0-J3 quadruple, and the large included the A0-G1-J2-K0 quadruple, which was interchanged with the A0-G1-J2-J3 to maximize $uv$ coverage.}
    \label{fig:uv}
\end{figure}

\subsection{MATISSE/VLTI}

We obtained high-angular-resolution interferometric data with MATISSE/VLTI in the $L$ (2.9 -- 4.1 \micron), $M$ (4.5 -- 4.9 \micron), and $N$ (8 -- 13 \micron) bands in ``imaging'' mode from 29 November, 2022 until 25 March, 2023 (ESO Program ID: 0110.C-4209(A), PI: F.~Lykou), while Z~CMa was in outburst (Fig.~\ref{fig:vislc}). In the imaging mode, all three array configurations (``small'', ``medium'', and ``large'') of the 1.8-m Auxiliary Telescopes (ATs) were utilized in order to obtain the best achievable $uv$-plane coverage (Table~\ref{tab:matlog}), and this is shown in Fig.~\ref{fig:uv}. Each hour of observation includes data acquisition for both science target and a calibrator in all three bands.

The MATISSE instrumental field of view (FOV) is that of the beam entering the instrument through a pinhole, and for the case of an AT with a given diameter, $D$ (1.8 m), it is $1.5 \lambda /D$ in the $LM$ band, and $2 \lambda /D$ in the $N$ band \citep{matisse}. Therefore, the FOV at 3.5 \micron\ and 10.5 \micron\ becomes 0.6\arcsec\ and 1.8\arcsec , respectively. As such, both stars in Z~CMa were observed simultaneously by MATISSE and it was not possible to define individual pointings.

The ``GRAVITY for MATISSE'' (GRA4MAT) mode \citep{gra4mat} was used for all observations, where the GRAVITY instrument -- the VLTI instrument operating in the $K$ band (2--2.4 \micron) -- is utilized as a fringe tracker to allow for better stabilization of the fringes of MATISSE, improving signal-to-noise (S/N) and calibration. Several spectral setups were chosen for these observations. For the imaging runs, we opted for the medium spectral resolution in the $L$ band (MED-L; $R\sim500$) and the low spectral resolution in the $N$ band (LOW-N; $R\sim30$).

Data reduction and analysis were performed with the MATISSE DRS pipeline ver. 1.7.5 and with the dedicated Python library {\tt mat\_tools}\footnote{\url{https://gitlab.oca.eu/MATISSE/tools}} for flux calibration and visualization. The basic concepts of the MATISSE data reduction have been introduced extensively in several other works \citep[e.g.,][]{varga2021,gamezrosas2022}; therefore, we refrain from repeating them here. Only one calibrator was used for both spectral bands for the imaging observations. Those were $\alpha$ CMa (A1V, diameter 6.09 mas) for the small array configuration and HD 48217 (K5III, diameter 2.60 mas) for the medium and large arrays. An assessment of the data quality and caveats is given in Appendix~\ref{sec:matqual}. 

Our data are complemented with guaranteed time observations (GTO) obtained on 23 January, 2021 and 15 November, 2021 during quiescence (Fig.~\ref{fig:vislc}) with the medium array configuration (ESO Program ID: 0106.C-0501(B) and 0108.C-0385(D); PI: B.~Lopez). The GTO runs (Table~\ref{tab:matlog}) contained experimental observations in the $L$ band with both medium and low spectral resolutions (LOW-L: $R\sim30$), and both the low and the high spectral resolution in the $N$ band (HIGH-N: $R\sim220$). The first GTO data were obtained in a CAL\_N-SCI-CAL\_L sequence, whereby a different calibrator was used for each spectral band, namely $\alpha$ CMa (A1V, diameter 6.09 mas)  as CAL\_N and HD58972 as CAL\_L (K3III, diameter 3.39 mas). It is worth noting that since MATISSE observations were obtained simultaneously in both bands, the science target's data can be reduced with both calibrators to check consistency. The second GTO observations were obtained with one calibrator for both bands (HD39853; K4III, diameter 2.23 mas).

\subsection{SpeX/IRTF}

We obtained long-slit spectra with the SpeX spectrograph \citep{rayner2003} on the Infrared Telescope Facility (IRTF) on Mauna Kea, Hawaii, on 04 November, 2023 (program ID: 2023B039, PI: P. \'Abrah\'am) while the system was still in outburst. We opted for the Echelle grating in short-wavelength mode (SXD), which covers the NIR wavelength region (0.7 -- 2.55 \micron). The achieved spectral resolution was $R\,\sim\,750$. The slit width of 0.8\arcsec\ covers both stars, and the spatial resolution is insufficient to recover individual stellar spectra.

HIP24555 (A0V) was chosen as the telluric standard. Although we did not obtain simultaneous photometry, we found that the telluric standard was sufficient for photometric calibration. Data reduction, corrections for telluric absorption, and flux calibration were performed using the {\sc Spextool} package \citep{vacca2003,cushing2004}.

The SpeX spectrum is rich in emission lines (red; Fig.~\ref{fig:specsed}), whose analysis will be the basis of a forthcoming paper. Here, we show that the photometric calibration of MATISSE agrees with SpeX, and we use both to build Z~CMa's SED in Fig.~\ref{fig:specsed}.

\section{Results}\label{sec:res}

The interferometric data -- that is, the correlated fluxes (Figs.~\ref{fig:fcorrL}-\ref{fig:fcorrN}), visibilities (Figs. \ref{fig:vis2L}-\ref{fig:vis2N}), and closure phases (Figs. \ref{fig:cpL}-\ref{fig:cpN}) -- show a sinusoidal modulation in several baselines in all three bands, indicating that the binary system is resolved. The amplitude of the sinusoidal modulation appears shallower in the $M$-band correlated fluxes and visibilities than in the $L$ and $N$ bands. In the following section, we provide initial estimates of the binary's separation and position angle through geometric model fitting of the visibilities and closure phases during quiescence. Image reconstruction results are shown in Sect.~\ref{imrecon}. The $N$-band correlated fluxes obtained with the small array show a deep absorption feature which we explore further in Sect. \ref{sec:radsil}.

We note the presence of several emission lines in the $L$- and $M$-band correlated fluxes from the small array observations (Figs.~\ref{fig:fcorrL} and \ref{fig:fcorrM}), which, as mentioned in the caveats in Appendix~\ref{sec:matqual}, belong to the hydrogen lines of the calibrator Sirius. Although these could be subtracted by averaging the emission around the line, we refrained from doing so here to indicate the adverse effect of this calibrator.

\begin{figure*}[!htbp]
    \centering
    \includegraphics[width=\linewidth]{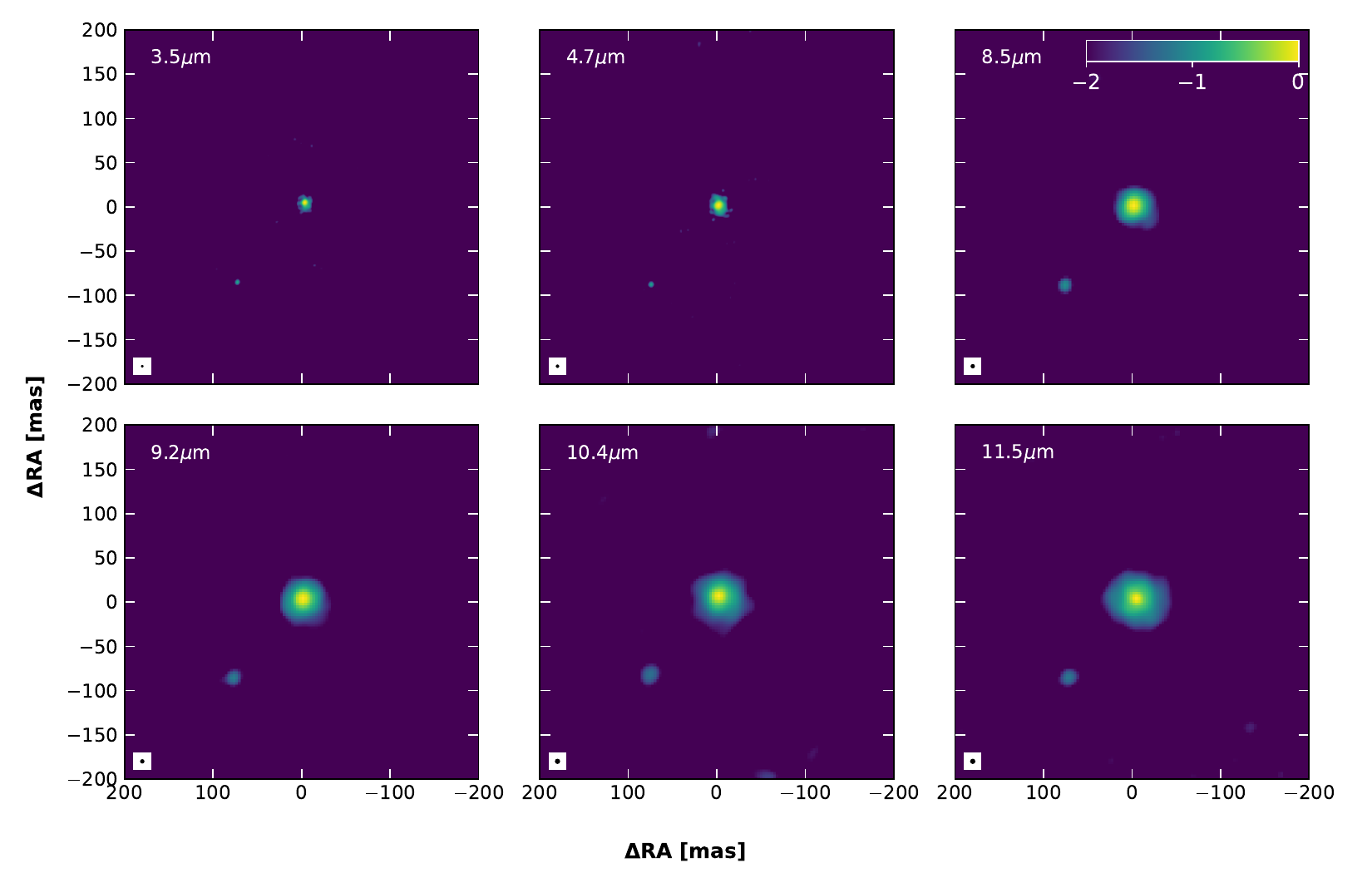}
    \caption{MiRA image reconstructions in $L$, $M$, and $N$ bands during the 2023 outburst. From left to right and top to bottom, these correspond to pseudo-continuum reconstructed images at 3.5, 4.7, 8.5, 9.2, 10.4, and 11.5 \micron. These are smoothed with a Gaussian beam with a FWHM similar to the image resolution, normalized to unity, and shown in a logarithmic stretch with a minimum value of $10^{-2}$ to suppress image reconstruction residuals below that level. Such residuals are the result of insufficient $uv$ coverage (Fig.~\ref{fig:uv}).  The SE component (FUor) becomes much fainter than the NW component (HBe) at longer wavelengths. The beam size is shown as a black circle at the bottom left corner of each panel. }
        \label{fig:mirarecon}
\end{figure*}

\begin{table*}[htbp]
    \centering
    \caption{{\tt LITpro} geometric model fitting results for quiescent state (id=1; Table~\ref{tab:matlog}).}
    \begin{tabular}{ccccccc}
    Method & $\theta_{\rm HBe}$ (mas) & $\theta_{\rm FUor}$ (mas) & $\Delta f$ & $\chi_{\rm red}^2$ & $\Delta\alpha$ (mas) & $\Delta\delta$ (mas) \\
    \hline\hline
    \multicolumn{7}{c}{$L$-band} \\
    Uniform disk & $16.1\pm0.03$ & $8.8\pm0.03$ & $1.5\pm0.5$ & 438.2 & $74.79\pm0.04$ & $-87.94\pm0.14$\\
    Gaussian FWHM & $10.8\pm0.2$ & $8.6\pm0.2$ & $11.5\pm0.3$ & 81.1 & $76.31\pm0.06$ & $-84.72\pm0.15$\\
    Binary star & - & - & $1.1\pm3.7$ & 27950  & $80.42\pm0.07$ & $-73.21\pm0.17$ \\
    \hline
    \multicolumn{7}{c}{$N$-band} \\
    Uniform disk & $44.3\pm0.1$ & $20.9\pm0.03$ & $1.3\pm0.1$ & 2.8 & $70\pm0.0$ & $-91.56\pm0.1$\\
    Gaussian FWHM & $30.2\pm0.1$ & $12.2\pm0.1$ & $1.6\pm0.1$ & 4.5 & $80.85\pm0.12$ & $-82.58\pm0.15$\\
    Binary star & - & - & $1.0\pm0.1$ & 2218 & $78.81\pm0.06$ & $-91.78\pm0.10$ \\
        \hline
    \end{tabular}
    \label{tab:litpro}
\end{table*}

\subsection{Geometric model fitting}\label{sec:geom}

Although this work focuses on the imaging results with MATISSE, we performed initial geometric model fits to the visibilities and closure phases of the GTO LOW-L and HIGH-N data obtained during quiescence (id: 1, Table~\ref{tab:matlog}) to derive approximate size estimates for the two sources, as well as relative astrometry, and compared these to literature values and the imaging results. 

We employed {\tt LITPro} \citep{litpro}, a geometric modeling tool that does not take into account the chromaticity of the data. For simplicity, we followed literature studies and fit centrosymmetric distributions \citep{ratzkaphd,monnier2005,millangabet2006,benisty2010} such as a uniform disk, a Gaussian distribution, and a binary in the form of two point sources, where the size of each source, the flux ratio, and the position of the companion are free parameters, while the Herbig star's position is fixed at the origin (0,0). 
The results of the {\tt LITPro} fitting process are shown in Table~\ref{tab:litpro}. These include the angular sizes of each disk ($\theta_{\rm HBe}$ and $\theta_{\rm FUor}$), their relative flux ratio $\Delta f$ (HBe to FUor), the relative position of the FUor companion with respect to the Herbig star ($\Delta \alpha$, $\Delta \delta$), and the goodness-of-fit $\chi_{\rm red}^2$.

Many of the fits are poor as the flux contributions and the size of the FUor were often highly correlated. This is more prominent when fitting two point sources. Since {\tt LITPro} fits the entire spectral range of the MATISSE $L$ and $N$ bands, and does not take into account either the chromaticity of each source or the probable temperature gradient in each disk, comparisons to the results from Sect.~\ref{sec:dis} and Table~\ref{tab:disks} could be tenuous. Attempts to improve such fitting in {\tt LITPro} by introducing background emission and/or incorporating a blackbody temperature for each disk failed to produce reliable results since the flux-ratio parameters were highly correlated and/or degenerate. Nevertheless, the geometric model fitting results can be interpreted as showing that the size of the HBe disk was larger than that of the FUor, irrespective of the geometric model during quiescence in 2021. For consistency, we derived the average relative astrometry from the three methods in the $L$ band, since the spatial distribution of the dust continuum is expected to affect the estimates in the $N$ band. As such, from Table~\ref{tab:litpro} we calculate $\left<\Delta\alpha\right>=77.17\pm0.03$ mas and $\left<\Delta\delta\right>=-81.96\pm0.09$ mas for the $L$ band (see Table~\ref{tab:relastro}). This result is included in our analysis in Sect.~\ref{sec:binary}.

From the above, we argue that the simple geometric fitting is convenient for obtaining initial parametric results on the source sizes and good estimates of the relative astrometry; however, it is not always reliable since it does not take into account the brightness distribution and intricate geometries (e.g., asymmetries) of each source. Since this work is focused on imaging data with MATISSE, we expect the reconstructed images to shed more light on the spatial distribution of both disks.

\subsection{Image reconstruction}\label{imrecon}

The image reconstruction concentrates on the MATISSE data obtained during the outburst phase (Fig. \ref{fig:vislc}). As mentioned in Sect.~\ref{sec:obs}, the MATISSE FOV is 0.6\arcsec\ and 1.8\arcsec\ at 3.5 \micron\ and 10.5 \micron , respectively. For consistency, we opted to use a common FOV and pixel scale for all reconstruction methods. Using the full FOV of MATISSE did not improve any of the initial image reconstruction tests, as it only increases the amount of residual noise in the maps while increasing computational time. Therefore, in all further image reconstruction the FOV was $400 \times 400$ mas with a pixel scale of 1 mas/pixel in the $LM$ band and 3 mas/pixel in the $N$ band. These pixel scales are used to achieve "super-resolution" (i.e., Nyquist sampling for each image), as that is estimated at the longest baseline obtained. 

Pseudo-continuum images were reconstructed in all bands, that is, for short bandwidths where no spectral lines are observed. These were at 3.4 -- 3.6 and 4.6 -- 4.8 \micron\ in the $L$ and $M$ bands, and 8.4 -- 8.6, 9.1 -- 9.3, 10.3 -- 10.5, and 11.4 -- 11.6 \micron\ in the $N$ band. Henceforth, the pseudo-continuum images are mentioned as the 3.5, 4.7, 8.5, 9.2, 10.4, and 11.5 \micron\ images, respectively. All image reconstructions are smoothed with a Gaussian of similar FWHM to the image resolution, that is, 1~mas in the $LM$ and 3~mas in the $N$ bands.

We used the online tool {\tt OImaging}\footnote{\url{https://www.jmmc.fr/english/tools/data-analysis/oimaging/}} to test two image-reconstruction methods (BSMEM and MiRA) for consistency and finally opted for the MiRA algorithm (see Appendix~\ref{sec:imreconmethods}). All array configurations with ids=3 to 23 and data quality ranked as A and B were used (see Table \ref{tab:matlog}). The reconstructed images are shown in Fig.~\ref{fig:mirarecon} and are ordered by increasing wavelength from left to right and top to bottom. All images are normalized to unity and shown in a logarithmic stretch with a minimum value of $10^{-2}$. This level is similar to the arbitrary flux-level minimum set during image reconstruction\footnote{Associating absolute flux calibration with image reconstruction methods is a contentious issue, since these methods rely only on visibilities and closure phases.}. Any structures seen in the maps below that limit are residual artifacts due to incomplete $uv$-plane coverage (see Appendix~\ref{sec:imreconmethods}; \citealt{planq2024} offers an extensive treatise on MATISSE image reconstruction). The theoretical MATISSE beam at each wavelength is shown in the bottom left corner of each panel.

The primary HBe is the obvious brighter and larger source in the middle of each panel in Fig.~\ref{fig:mirarecon}. The FUor companion is located as expected toward the southeast. All pseudo-continuum reconstructed images in Fig.\ref{fig:mirarecon} are shown at the same FOV to reveal the changes in the angular size of the sources. This is more prominent for the HBe primary. On the other hand, the FUor's size remained nearly constant at all wavelengths.

The separation of the Z~CMa binary was calculated from the reconstructed images in Fig.~\ref{fig:mirarecon}. Each position was estimated by finding the location of the brightest peaks in each map (in all bands) with the DAOFIND algorithm of the {\tt photutils} package\footnote{\url{https://photutils.readthedocs.io}} and then averaging over all maps since the observations were obtained simultaneously in all bands. Here, we did not take into account the temporal spread of the observations over four months (Table~\ref{tab:matlog}), since this is a wide binary with a rather large period. The positional errors are the standard deviations of the measured locations. Our MATISSE measurements result in a separation $117.88\pm 0.73$ mas and at a position angle of $\rm 139.16^o \pm 0.29^o$ east of north. This epoch will be included in Sect. \ref{sec:binary}.

Apart from the relative astrometry, the DAOFIND algorithm also calculated  fluxes for each star, that is, the central bright region in each image. Since the pseudo-continuum reconstructed images are not calibrated with respect to absolute photometry, and considering that this is a nontrivial matter in image reconstruction, these fluxes are only relative. As such, we estimated that the relative-flux ratios $\Delta f$ at the nearest decimal of the HBe with respect to the FUor, and these can be found in Table~\ref{tab:relflux}. Here, the uncertainties reflect the range of values for apertures of 1-3 pixels wide.

\begin{table}[bhtp]
    \centering
    \caption{Relative flux ratios of HBe with regard to FUor.}
    \begin{tabular}{cc}    
        \hline
        $\lambda\,(\mu\rm m)$ & $\Delta f$ \\ \hline
        3.5 & $2.8\pm0.2$ \\
        4.7 & $2.4\pm0.2$ \\
        8.5 & $6.0\pm0.2$ \\
        9.2 & $9.5\pm0.3$ \\
        10.4 & $24.4\pm10.0$ \\
        11.5 & $27.0\pm10.0$ \\
        \hline
    \end{tabular}
    \label{tab:relflux}
\end{table}

It appears that in the $N$ band the FUor companion is more than ten times fainter than the HBe primary. As a reminder, the MATISSE imaging data were obtained while the HBe star was in outburst. The brightness ratio between the FUor and HBe star has been studied before. In the infrared and up to the $L'$ band, \citet{hinkley2013} and \citet{bonnefoy2017} have shown that the flux ratio of the HBe with respect to the FUor increases at longer wavelengths, especially while the primary is in outburst. It is worth mentioning that the binary cannot be resolved with conventional AO-assisted imaging beyond the $L$ band. \citet{ratzkaphd} estimated an HBe/FUor flux ratio $\approx6.3^{+2.8}_{-1.5}$ in the $N$ band by fitting the visibilities of MIDI data obtained while the system experienced a short-term burst in late 2004. In that respect, our results are consistent with literature studies of the (out)burst phase. Moreover, \citet{hinkley2013} reported a continuous decay in the flux of the FUor companion with respect to historical values. Considering that their observations were obtained approximately 15 years ago (at the time of writing), it is worth revisiting the system to explore whether the FUor continues its dimming and if it has returned to quiescence. 

\begin{figure}
    \centering
    \includegraphics[width=\columnwidth]{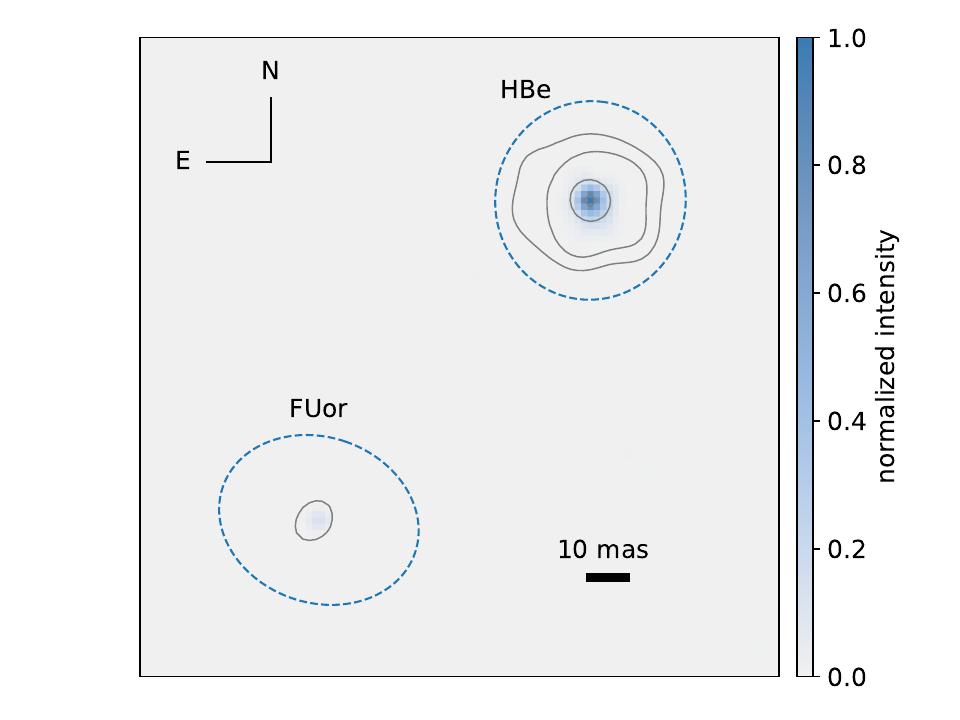}
    \caption{Comparison between ALMA and MATISSE. The normalized pseudo-continuum reconstructed image at 3.5 \micron\ is shown in blue. Overlaid are contours (gray lines) at 5, 10, 50, and 90\% of the peak of the normalized 9.5 \micron\ reconstructed image. These are compared to the ALMA 1.3-mm continuum disk sizes (blue dashed-line ellipsoids) of \citet{dong2022}. }
    \label{fig:alma_mat}
\end{figure}

Figure~\ref{fig:alma_mat} shows a comparison of the protoplanetary disk sizes between  MATISSE and ALMA \citep{dong2022}. The underlying reconstructed image (in blue) corresponds to the 3.5 \micron\ pseudo-continuum, while the overlaid gray contours correspond to the 9.5 \micron\ image reconstruction.  Both have been normalized to the maximum peak brightness (here, the Herbig star), and the contour levels are at 5\%, 10\%, 50\%, and 90\% of the peak. This comparison shows that the majority of the MIR flux arises from the same region at 3.5 and 9.5 \micron\ for the Herbig disk (size$\le$15~mas), while the disk may extend up to $\sim$40~mas in size (upper limit) at lower flux levels. The FUor disk's size remains somewhat constant in both wavelengths (size $\ll$15~mas), although it is much fainter compared to the HBe at 9.5 \micron\ (5\% contour level). Also overlaid in Fig.~\ref{fig:alma_mat} are two ellipsoids (blue dashed lines), which correspond to the maximum deconvolved disk sizes (Gaussian FWHM; see Table~\ref{tab:disks}) from the ALMA 1.3-mm continuum in \citet{dong2022}. For reference, the ALMA beam was $78\times46$ mas and PA 66\degr\, and the size of both sources prior to deconvolution were similar to the beam size, while \citet{dong2022} caution that ``extended dust emission may affect their fitting process.'' The latter could explain the discrepancy in disk inclination between 1.3-mm and 9-mm data (see Table~\ref{tab:disks}). The optically thin and cooler outer regions of protoplanetary disks that emit in the (sub)millimeter regime are expected to be significantly larger than the optically thick inner regions that are emitting in the NIR and MIR \citep[see, e.g.,][]{dullemond2010}.

\subsection{Radial variation of the silicate feature and optical depth analysis}\label{sec:radsil}

The MATISSE $N$-band correlated fluxes\footnote{Also known as correlated spectra; henceforth, both expressions used in this work are interchangeable.} are shown in Fig.~\ref{fig:fcorrN}. Each panel corresponds to a separate array configuration to those in Table~\ref{tab:matlog} for the data obtained during the 2022/3 outburst. As a reminder, these are the so-called imaging data not to be confused with the GTO data obtained during quiescence. 

For the medium and the two large arrays (Table~\ref{tab:matlog}) in Fig.~\ref{fig:fcorrN}, the $N$-band correlated spectra are dominated by three factors: (a) the sinusoidal modulation of the binary, (b) the effects of the terrestrial atmosphere (atmospheric seeing, wind speed, and coherence time), and (c) the rather noisy, low S/N for certain baselines. For the second case, the effect is increased for fainter calibrators such as the one used in these arrays. For the third case, these baselines are usually the combination of AT2-AT3 telescopes, that is, the shortest baselines for each array, and could be related to the known, occasionally poor performance of AT3. Considering these uncertainties, we excluded the correlated spectra obtained with these arrays from our analysis of the silicate feature. 

The small-array correlated spectra (Fig.~\ref{fig:fcorrN}) show the silicate feature in absorption. The strength (flux level) of the feature is decreasing with increasing baseline length, that is, from 10.4 m to 33.9 m (or else at spatial scales from approximately 32 to 104 mas). We note that despite our concerns about using a large calibrator, such as Sirius (see Appendix~\ref{sec:matqual}), the overall shape and strength of the feature are consistent with MIDI/VLTI observations obtained with different calibrators but with similar baselines \citep{varga2018}. Therefore, the radial variation of the strength of the silicate absorption feature with regard to baseline length is real. 

The binary can be resolved even with the small array data in the $N$ band, although the disks are marginally resolved (size $<$~10~mas) and no other spatial information can be derived for them. Therefore, it is possible that the $N$-band correlated spectra also contain the sinusoidal modulation of the binary's signal, which could slightly shift the minimum of the silicate feature. However, this appears to be stationary taking into account its proximity to the 9.4 -- 9.9 \micron\ ozone feature that affects that part of the spectrum.

This radial variation of the correlated spectra may be related to changes in the optical depth of dusty material along the line of sight; that is, the correlated fluxes are attenuated by a factor of $\propto e^{-\tau_{\lambda}}$ \citep[e.g.,][]{lykou2024}. The interstellar extinction should also be taken into account \citep[e.g.,][]{quanz2007} before considering estimating the circumstellar extinction, and to that end we dereddened each correlated spectrum using $A_V=2.5$ (see Appendix~\ref{ism}). Following that, we explored this attenuation using the methods of \citet{vanboekel2003} and \citet{kemper2004}, whereby we fit a linear continuum $F_{\rm c}$ between 8 and 12.9 \micron\ for each correlated spectrum $F_{\rm corr}$, and we extracted the optical depth as $\tau_{\lambda} = -\ln({F_{\rm corr}/F_{\rm c}})$. The calculated optical depths per epoch and per baseline are shown in Fig.~\ref{fig:alltau}. We note that $\tau_{\rm \lambda} <0.3$ overall. We discuss our analysis further in Sect.~\ref{sec:disktau}.

\section{Discussion}\label{sec:dis}

\begin{table*}[htbp]
    \centering
    \caption{Z~CMa angular sizes from literature and from MATISSE reconstructed images.}
    \begin{tabular}{cccccccc}
    \hline
    Reference & Epoch & State & $\lambda$ & major (mas) & minor (mas) & PA (\degr) & inclination(\degr) \\ \hline
    \multicolumn{8}{c}{Z CMa NW (Herbig Be)} \\
    \citet{monnier2005}          & 2004 & Q & 2.2 \micron\ & $3.95\pm0.24$ & -- & -- & -- \\
    \citet{benisty2010}          & 2008 & O & 2.2 \micron\ & $\sim$3.4 & -- & -- & -- \\
    \citet{dong2022}             & 2016/7 & B & 1.3 mm & $23\pm6.8$ & $22\pm9.1$ & $82\pm86$ & $16.9\pm95.5$  \\
    \citet{dong2022}             & 2016 & B & 9 mm & $37\pm7.7$ & $18\pm13$ & $80\pm49$ & $60.9\pm23.9$  \\
    \citet{varga2018}           &  model & B & $N$ & $\sim$7.5 & -- & -- & -- \\
    this work                    & 2022/3 & O & 3.5 \micron\ & $4.8\pm0.05$ & $4.1\pm0.04$ & $165.7\pm3.7$ & $30.3\pm2.0$   \\
    this work                    & 2022/3 & O & 4.7 \micron\ & $7.4\pm0.1$ & $6.4\pm0.1$ & $165.9\pm2.6$ & $30.2\pm1.9$   \\
    this work                    & 2022/3 & O & 8.5 \micron\ & $13.7\pm0.2$ & $11.9\pm0.14$ & $166.6\pm3.5$ & $29.3\pm2.4$   \\
    this work                    & 2022/3 & O & 9.2 \micron\ & $13.9\pm0.2$ & $12.5\pm0.5$ & $159.2\pm4.7$ & $25.7\pm2.9$   \\
    this work                    & 2022/3 & O & 10.4 \micron\ & $13.6\pm0.2$ & $12.8\pm0.2$ & $148.6\pm9.3$ & $19.6\pm4.4$   \\
    this work                    & 2022/3 & O & 11.3 \micron\ & $14.6\pm0.3$ & $13.3\pm0.3$ & $133.3\pm8.8$ & $24.6\pm5.3$   \\
    \hline
    \multicolumn{8}{c}{Z CMa SE (FUor)} \\
    \citet{millangabet2006}      & 2004 & Q & 2.2 \micron\ & $3.94\pm0.24$ & -- & -- & -- \\
    \citet{dong2022}             & 2016/7 & B & 1.3 mm & $26\pm5.8$ & $20\pm5.9$ & $71\pm61$ & $39.7\pm25.5$ \\
    \citet{dong2022}             & 2016 & B & 9 mm & $59\pm9.8$ & $33\pm12$ & $38\pm23$ & $55.9\pm15.4$ \\
    this work                    & 2022/3 & O & 3.5 \micron\ & $3.1\pm0.1$ & $2.3\pm0.05$ & $157.0\pm3.2$ & $41.5\pm2.9$   \\
    this work                    & 2022/3 & O & 4.7 \micron\ & $3.8\pm0.1$ & $3.3\pm0.1$ & $166.9\pm7.0$ & $28.8\pm5.0$   \\
    this work                    & 2022/3 & O & 8.5 \micron\ & $9.4\pm0.4$ & $7.4\pm0.3$ & $162.9\pm6.8$ & $38.4\pm6.1$   \\
    this work                    & 2022/3 & O & 9.2 \micron\ & $11.7\pm0.4$ & $9.6\pm0.3$ & $151.1\pm6.3$ & $24.6\pm5.1$   \\
    this work                    & 2022/3 & O & 10.4 \micron\ & $15.8\pm0.3$ & $12.8\pm0.2$ & $149.7\pm3.1$ & $36.0\pm2.6$   \\
    this work                    & 2022/3 & O & 11.3 \micron\ & $13.7\pm0.3$ & $11.6\pm0.2$ & $127.6\pm4.6$ & $32.0\pm3.5$   \\
    \hline
    \multicolumn{8}{c}{unresolved components} \\
    \citet{monnier2009}          & 2005 & O & 10.7 \micron\ & $68\pm2$ & $41\pm6$ & $137\pm5$ & -- \\
    \hline
    \end{tabular}
    \tablefoot{Disk sizes correspond to Gaussian FWHM. In this work, these correspond to more than 50\% of the peak in the reconstructed images. For (sub)millimeter and radio observations, these are estimates from deconvolved continuum images, while for \citet{monnier2005} this is the diameter of a uniform ring. Column 3 describes the variability state of the Herbig star during each observation: outburst (O), quiescence (Q), and observed during both states (B). }
    \label{tab:disks}
\end{table*}

\subsection{Protoplanetary disks}

Earlier attempts to image the Z~CMa system in the MIR with AO-assisted telescopes failed to resolve the binary. \citet{malbet1993} suggested a circumbinary disk with a size of 0\farcs38 $\pm$ 0\farcs06 and a PA 161\degr $\pm$ 8\degr in the $L'$ (3.74~\micron ) and $M$ (4.75~\micron ) bands. On the other hand, \citet{monnier2009} could not fully resolve the binary in the $N$ band, but they only found an extension toward the FUor location. They measured a source size of $(68\pm2)\times(41\pm6)$ mas at the presumed location of the Herbig star with a PA at 137\degr$\pm$5\degr\ and an over-resolved (halo) component at $(2\pm1$)\% of the total flux. The MATISSE images do not reveal any circumbinary warm dust emission.

Table~\ref{tab:disks} lists the protoplanetary disk sizes corresponding to a 2D Gaussian fit to each disk in the MATISSE image reconstructions for the brightest regions (that is $\ge 50\%$ of the peak). Based on their apparent separation on the sky at less than 120~mas, the radius of the disks ought to be truncated to less than half the separation; that is, $<60$~mas, or $<70$~au, at our adopted distance. Our findings for each individual disk are discussed further below. 

Apart from this work, there exist only a handful of interferometric studies of the Z~CMa system from the last 20 years. Most of them observed the system while the Herbig star was in quiescence \citep{monnier2005, millangabet2006, dong2022}, while two studies observed it during an (out)burst \citep{monnier2009, benisty2010}. The source size estimates from the literature are also listed in Table~\ref{tab:disks}. Also listed are the protoplanetary disks' orientations where available. For the \citet{dong2022} deconvolved 1.3 and 9 mm continuum disk sizes, we calculated the inclination as $i=\arccos{ (a_{\rm minor}/a_{\rm major})}$, and we note the large uncertainties based on error propagation.

\subsubsection{The FUor companion}

In the early 1990s, the FUor companion was found to be the brightest of the pair in the visual \citep{barth1994}. More recent studies suggest that the FUor is slowly dimming both in the visual and in the infrared \citep{hinkley2013, bonnefoy2017, sag2020}. The diameter of its hot, innermost region was found to be less than 4~mas in the $K$ band \citep{millangabet2006}, which translates to $\leq2.2$~au radius at our adopted distance of 1125 pc.

Our analysis of the MATISSE data shows that the MIR emitting regions of the FUor disk is resolved in all bands. At shorter wavelengths, that is, in the $L$ and $M$ bands, the majority of the emission ($\ge50\%$ peak) rises from a region with a size (i.e., diameter) of $\le 4$~mas, while at lower levels (1\% peak) it extends to $<9$~mas. In the $N$ band, the brightest disk region extends up to $\sim 15$~mas ($\ge50\%$ peak) and $<20$~mas at lower levels. When compared to the geometric model fitting from the quiescent epoch data, the angular sizes at lower flux levels agree relatively well (Sect.~\ref{sec:geom}). This is not surprising since it is the Herbig star that went into outburst, and the size of the FUor disk should not have been affected.

Overall, the innermost disk region emitting in the MIR has a radius of $<5$~au, which is consistent with the $K$-band size mentioned above, confirming that the inner disk is compact. Similar results have been found for other FUor systems \citep{bourdarot, lykou2022, lykou2024}. The cooler, outer parts of the FUor disk as measured by ALMA, even though marginally resolved at 1.3 mm, are indeed wider than those of MATISSE with a radius of $\leq$30~au.

Our estimates for the disk orientation from the MATISSE image reconstructions have large uncertainties (Table~~\ref{tab:disks}). The mean position angle of the disk is 152.5\degr$\pm$12.7\degr\ and the mean inclination is 33.5\degr$\pm$5.7\degr . The uncertainties are the standard deviations of the individual estimates. Given the extremely large uncertainties in the disk PAs and inclinations calculated from the millimeter continuum maps of \citet{dong2022}, we cannot confirm a misalignment between the warm inner disk and its cooler exterior. On the other hand, the disk PA agrees rather well with the orientation of the FUor's jet; that is, approximately 235\degr\ \citep{whelan2010,antoniucci2016}.

\subsubsection{The Herbig primary}

The HBe star is responsible for the EXor-like (out)bursts of the last 20 years \citep[Fig.~\ref{fig:vislc}, and][]{sag2020}. Previous interferometric studies in the NIR ($K$-band), which were obtained during quiescent and outburst phases, have found that the diameter of its hot, inner disk region is smaller than $\le 4$ mas \citep{monnier2005,millangabet2006,benisty2010}. In fact, the diameter appears to have remained relatively constant irrespective of the variability phase. MATISSE can marginally resolve that inner disk region in the MIR. Mid-infrared observations with MIDI/VLTI could resolve the pair of stars -- but no spatial information could be extracted for each disk -- and a binary model fit suggested the Herbig star was brighter with a flux ratio of $0.16\pm0.05$ at $\sim$10 \micron\ \citep{ratzkaphd}. However, an additional extended symmetric component with a $\theta_{\rm FWHM}\sim17$~mas had to be included in the models to reproduce the observed visibilities. That component could represent the regions probed with MATISSE well. \citet{varga2018} modeled the $N$-band MIDI data of the HBe disk with a temperature gradient model ($q\sim 0.87$) and a disk's inner rim at the dust sublimation radius estimated at $\sim$3.9~au for a luminosity\footnote{\citet{varga2018} presume that the luminosity originates from the outbursting accretion disk instead of the stellar photosphere. } of $\sim 3200$~L$_\odot$ and a dust sublimation temperature of $T_{\rm sub} = 1500$~K. The disk's half-flux radius was at $\sim$10.3~au. At their adopted distance of 1.05 kpc, the dust sublimation diameter from the temperature gradient model is $\sim$7.5~mas, which is almost double that of the $K$-band diameter estimates for a similar distance \citep{monnier2005,benisty2010}. The $N$-band half-flux diameter is $\sim$20~mas.

The MATISSE reconstructed images show that the majority of the emission ($\ge 50\%$ peak) arises from a region with a diameter $<5$~mas in the $L$ band and $<15$~mas in the $N$ band. This would translate to disk radii $<3$ and $<9$~au, respectively. The region seen in the $L$ band could correspond to dust continuum from the inner rim of the disk. Although the true luminosity of the HBe is not known, we can assume 5000~L$_\odot$ as a lower limit based on the SED fitting (Fig.~\ref{fig:specsed}). The typical sublimation temperature for silicate grains is 1500~K. Therefore, the dust sublimation radius at the given luminosity becomes $R_{\rm sub}\approx 5$~au. This radius can extend as far as 7~au when using the accretion luminosity calculated by \citet{fairlamb2015}, which was a factor of two larger (10\,000~L$_\odot$). It is therefore possible that the $L$-band emission arises from inside the theoretical dust sublimation temperature. On the other hand, the $N$-band region where dust emission dominates is more consistent with the latter case when assuming a higher luminosity.

If the discrepancy among the region detected at 2 \micron, the region detected in the $L$ band, the dust sublimation radius from the temperature-gradient model, and the above-mentioned theoretical $R_{\rm sub}$ estimates is real, then one could argue for a potential cavity or gap in the disk and whether that could be detected in the MIR \citep[e.g.,][]{maaskant2013, menu2015}. However, the dereddened SED of the 2023 outburst shows NIR excess, even for higher values of $A_V$, and it is relatively flat with a MIR excess (Fig.~\ref{fig:specsed}). These suggest a continuous disk. Unlike in \citet{hinkley2013}, which provided photometry for each star, the SpeX spectra and the averaged archival photometry used in this work originate from the entire system. However, we presume that the optical and NIR fluxes originate from the Herbig star that is in outburst. Overall, the current SED agrees well with the scenario that Z~CMa belongs to Group Ib of flared disks \citep{meeus2001}.

At lower levels, that is, down to 1\% of the peak, the diameter of the HBe disk is $\le 38$~mas at 10.4 \micron, and $\leq 48$~mas at 11.5 \micron. These can be compared to the estimates of \citet{monnier2009} at 10.7 \micron\ with a source size $68\times41$ mas at the presumed location of the Herbig star. Although that dataset was also obtained at an epoch when the Herbig star was in outburst, it is more likely that the difference in sizes is related to the different observing methods\footnote{The system was not fully resolved but implies an extension toward the FUor companion.} and not the stellar variability. The above-mentioned extent of the disk at those lower flux levels beyond 9 \micron\ would suggest that the warm inner HBe disk is of similar size or slightly larger that the cooler, optically thin dust emission detected at longer wavelengths by ALMA and JVLA \citep{dong2022}. This discrepancy might be explained if the MIR-emitting disk regions were significantly brighter and therefore appear larger during the 2023 outburst, as opposed to the cooler outer region that was imaged during quiescence (ALMA+JVLA; Fig.~\ref{fig:vislc}). However, the MIR spectrum of Z~CMa has not changed much overall within the last 20 years \citep[e.g.,][]{polomski2005,varga2018} within the 20\% flux uncertainties of MATISSE. Since this size discrepancy would violate the typical model for protoplanetary disks \citep{dullemond2010}, we believe that the millimeter size estimates are either underestimated if the disks were marginally resolved by ALMA and JVLA, or the millimeter-emitting region is more optically thick than expected. The latter might be the case if the spectral index of the HBe disk is $\alpha\le 2$ \citep{baobab2019}. The discrepancies between the MIR- and millimeter-emitting regions, as well as that of the inner disk rim and the dust sublimation temperature, could be explored with a radiative transfer model for the disk.

\begin{figure}
    \centering
    \includegraphics[width=0.9\columnwidth]{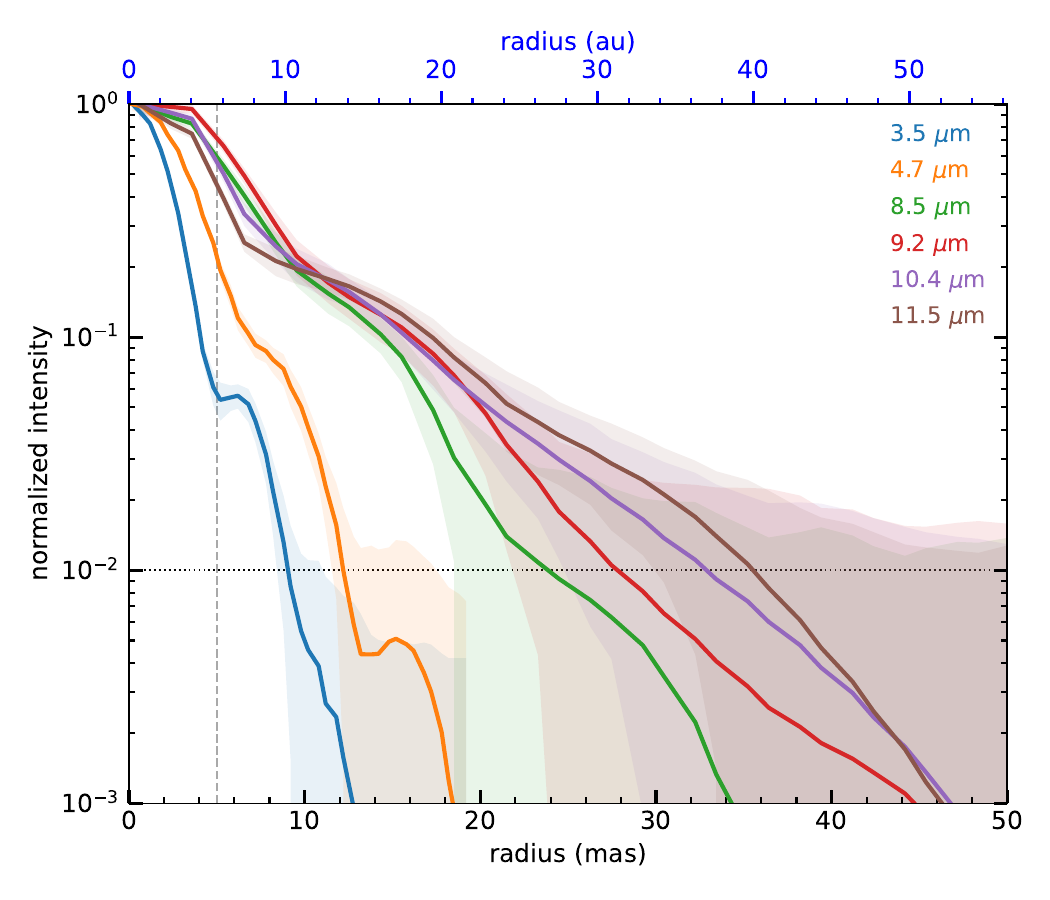}
    \caption{Normalized radial profiles of HBe disk from reconstructed images in Fig.~\ref{fig:mirarecon}. The horizontal dotted line demarcates the relative flux threshold set for the image reconstructions. The vertical dashed line demarcates the location of the first plateau at 3.5~\micron\  and, consequently, the region from which the majority of the emission originates at all wavelengths.}
    \label{fig:radprof}
\end{figure}

Normalized radial profiles were created from each reconstructed map in Fig.~\ref{fig:mirarecon} within a 60~mas radius of the peak-emission pixel, which we have to presume is the unresolved protostar. These radial profiles are shown in Fig.~\ref{fig:radprof}. We set a threshold at 1\% of the peak as applied to the image reconstructions. At a first glance, the radial profiles indicate a continuous disk whose brightness diminishes toward larger radii, as would be expected from a temperature gradient. This occurs inside a maximum radius of 50~au, congruent with the expected truncation radius from the binary's separation. The majority of the MIR emission originates within a 5~mas radius, that is, $<$6~au (vertical dotted line). This is the location of the first plateau in the 3.5 \micron\ reconstructed map. The inner disk region emitting at shorter wavelengths (3.5 and 4.7 \micron) is contained within a radius of $<$10~mas at its farthest extent (otherwise, $\le$11~au). The outer-disk radius can extend up to 36~mas in the $N$ band (i.e., $\le$40~au). The $N$-band profiles show a distinct shoulder at radii of approximately 8 -- 13~mas; that is, at the farthest extent of the inner-most region. Regarding the Herbig star's variability, it is difficult to ascertain real changes in the disk sizes between quiescence and outburst based on the simplicity of the geometric model fitting (Sect.~\ref{sec:geom}, Tables~\ref{tab:litpro}).

The MATISSE images suggest an average PA at 156.5\degr$\pm$~12.1\degr\ and inclination at 26.6\degr$\pm$~3.8\degr\ for the HBe disk (Table~\ref{tab:disks}). As is the case with its FUor companion, this orientation agrees well with the derived PA for the HBe jet \citep[245\degr ; ][]{whelan2010,antoniucci2016}. Moreover, the orientations of both disks agree well with each other within the uncertainties. This could suggest that the disks have not been perturbed by a tertiary companion. Once again, we do not find clear evidence for a misalignment between the warm inner HBe disk detected by MATISSE and its cooler outer counterpart viewed by ALMA and JVLA due to the large uncertainties in determining the orientation of the latter.

Based on all of the above, we do not find any clear evidence for substructures such as gaps in the HBe disk, and we do not find any evidence of misalignments either. The current observations might not have had sufficient resolution to reveal such structures, but perhaps this could be revisited in the future with the new 200-m baseline of the VLTI, while the same can be argued for future ALMA observations that could explore spatially resolved kinematics for each disk. Nevertheless, such observations will still be limited by the large distance to Z~CMa.

\subsection{Dust distribution }\label{sec:disktau}

Near-infrared polarimetric imaging studies of Z~CMa \citep{canovas2012,hinkley2013,canovas2015,liu2016} suggest that the Herbig star is enshrouded in a dusty cocoon. This could potentially be a portion of material spanning hundreds of astronomical units within the overall outflow \citep[e.g.,][]{zurlo}. Considering the wavelengths probed, this is due to dust scattering from small-sized grains distributed uniformly over the system, since none of these studies could resolve the binary behind coronagraphic masks. \citet{canovas2012} modeled the polarization seen in this extended (size $>$1\arcsec) cocoon with a spherical shell around the Herbig star extending to 70~au, with an inner hole at a radius of 20~au, which in turn is embedded in a large-scale, infalling molecular envelope. This morphology could allow for the emanation of a jet, a bipolar outflow, and/or disk winds from the HBe star \citep{poetzel1989, benisty2010}.

At our adopted distance, the cocoon's dimensions could be translated to angular sizes of $\sim$124~mas for the shell and $\sim$35~mas for the inner hole. Both scales are feasible for MATISSE, and here we explore whether the structures detected in the MIR could be related to the alleged cocoon seen in scattered light through our analysis of the radial variation of the silicate feature (Sect.~\ref{sec:radsil}).

\begin{figure}
    \centering
    \includegraphics[width=0.75\columnwidth]{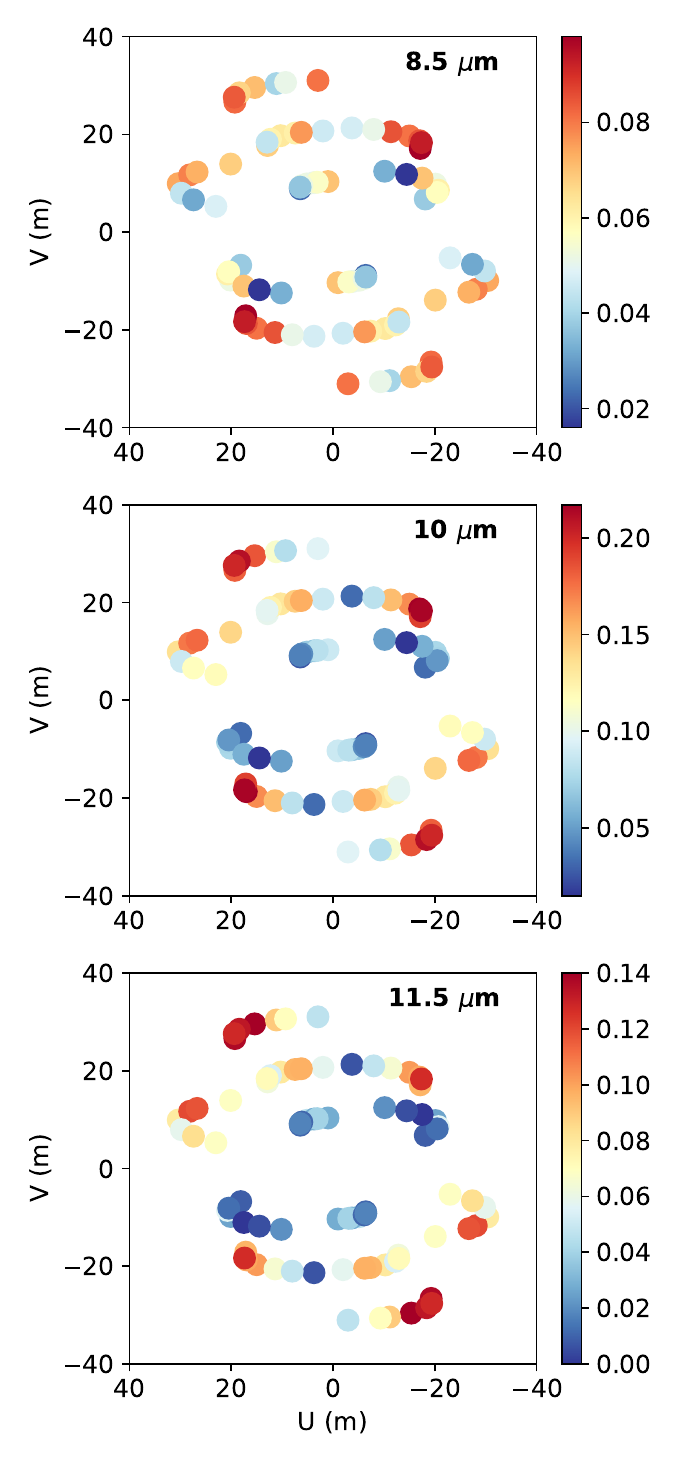}
    \caption{Optical-depth $uv$ maps of Z~CMa. These correspond, from top to bottom, to 8.5, 10, and 11.5 \micron, and the $uv$ coverage is that of the small array (see also Fig.~\ref{fig:uv}). The optical depth varies per wavelength, as indicated by each individual color bar, as well as per spatial scale probed by MATISSE (see Sects.~\ref{sec:radsil} and \ref{sec:disktau}). }    \label{fig:taumap}
\end{figure}

Despite the limited $uv$ coverage of MATISSE's small array data (see Fig.~\ref{fig:uv}), we explored the distribution of dust by mapping the optical depth at three different wavelengths, namely 8.5, 10, and 11.5 \micron. The optical depth values were extracted at these wavelengths for each baseline or for each $uv$ point (Fig.~\ref{fig:alltau}). We therefore re-imaged the optical depth distribution with respect to each $uv$ point. These $uv$ maps are shown in Fig.~\ref{fig:taumap}, and from top to bottom they correspond to 8.5, 10, and 11.5 \micron, while the strength of the optical depth is indicated in each individual color bar. 

These $uv$ maps show that the dust is not distributed uniformly over the source, but there is a preferential distribution: smaller $\tau_{\rm \lambda}$ at large spatial scales (i.e., toward the center of each map), and larger $\tau_{\rm \lambda}$ at small spatial scales (i.e., toward the outer regions of each map). Therefore, this nonuniform distribution is found on spatial scales of approximately 30 -- 100~mas or within an area with a size of 35 -- 110~au. Conversely, the separation of the binary is $>100$~au, while, as we showed earlier, the FUor companion is $>10$x fainter than the Herbig star in the MIR. 

\begin{figure}
    \centering
        \includegraphics[width=0.75\columnwidth]{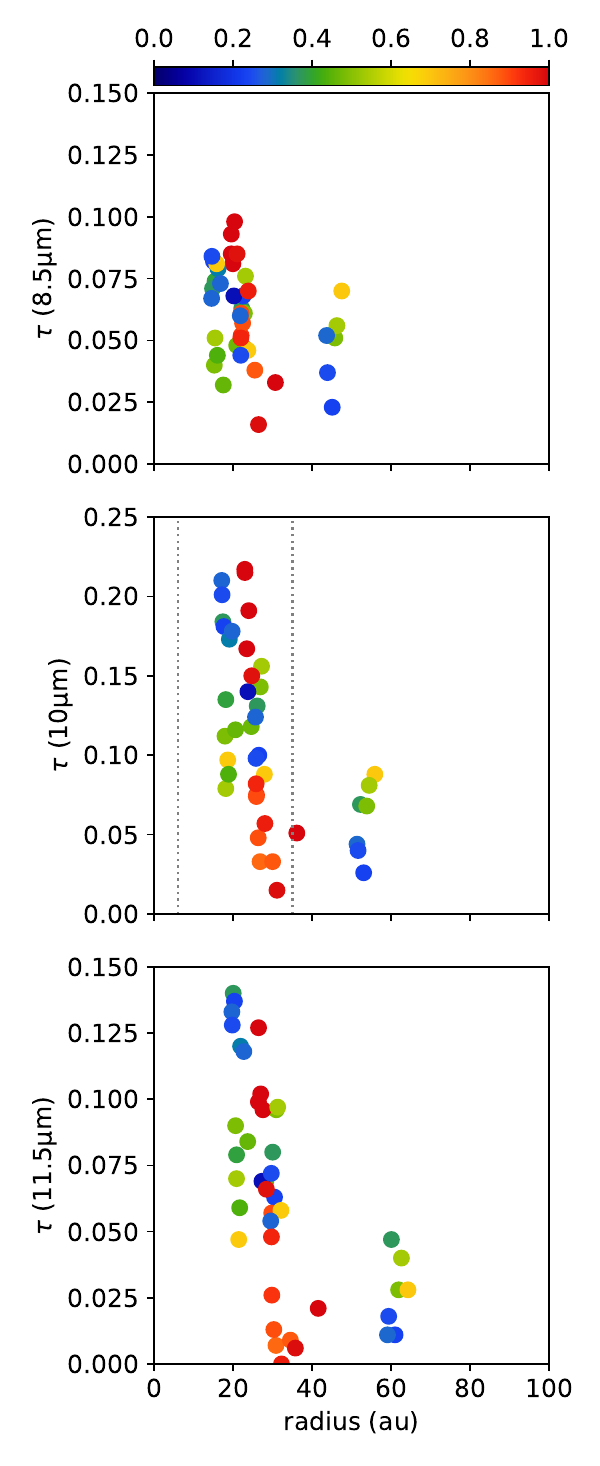}
    \caption{Optical depth variation at different physical radii. Three wavelengths are explored in the panels from top to bottom at 8.5, 10, and 11.5 \micron, respectively. The radius is a $1/2$ 
    spatial scale probed by the interferometer (see Sect. \ref{sec:disktau}) converted to physical units at the adopted distance. The optical depth is color-coded with regard to our parameter, $\eta$ (see Sect.~\ref{sec:disktau}). The vertical gray dotted lines in the middle panel mark the upper limit of the radial extent of the HBe disk in the 10.4 \micron\ image (Fig.~\ref{fig:mirarecon}) at 50\% and 1\% of the normalized peak. }
    \label{fig:taurad}
\end{figure}

The nonuniform distribution is more clearly illustrated in Fig.~\ref{fig:taurad}, where the $uv$ coverage of Fig.~\ref{fig:taumap} has been converted to angular radii ($\lambda /4B_{\rm \lambda}$), and these were in turn translated to physical radii for our adopted distance. The optical depth for each wavelength is color-coded with respect to the parameter $\eta$, which we define as:
\begin{equation}
    \eta = \lvert \cos\left(B_{\rm PA} - \theta\right) \rvert,
\end{equation}
where $B_{\rm PA}$ is the baseline PA and $\theta$ in the PA of the binary's orbit estimated from MATISSE data (Sect.~\ref{imrecon}). For baselines parallel to the binary's orbit, $\eta$=1, while for those perpendicular to it, $\eta$=0. However, the optical depth does not seem to be dependent on the baseline PA. 

As shown in the top panel of Fig.~\ref{fig:taurad}, the optical depth at 8.5 \micron\ is relatively flat, with $\tau_{\rm 8.5\mu m} <0.1$. At longer wavelengths (middle and bottom panels), the optical depth varies with respective physical radius. That is, the highest optical depth resides at radii of $< 40$~au. This is directly comparable to the extent of the HBe disk at 10.4 \micron\ (gray dotted lines) located at approximately 7 and 35~au (upper limits) at 50\% and 1\% of the normalized peak, respectively. Therefore, we suggest that the variation of the silicate feature is related to the Herbig star's protoplanetary disk and possibly material around it, while optically thin material lies beyond 40~au and could extend up to the presumed truncation radius of $\approx 65$~au.

Previous studies \citep{cohen1980,cohen1985,polomski2005,schutz2005,dodu2020} have explored the silicate absorption feature using MIR long-slit spectroscopy and spectro-photometry. Their modeling of the feature with different dust mixtures, primarily silicates, and assigning black-body temperatures for the absorbing material, suggest optical depths $\tau_{\rm \lambda} \approx$ 0.2 -- 0.5 and dust temperatures of 300 -- 500~K. As such, our estimates for the optical depth agree with literature values, although it is worth noting that they were constrained to radii of $\geq$20~au, and a moderate interstellar extinction correction has already been applied here. The optical depth would be much higher on smaller spatial scales near the protostar.

Several of the studies mentioned above suggested that the absorption feature originates from the FUor companion. The MATISSE images have shown that the FUor companion is indeed $>$10x fainter than the HBe primary. For the silicate feature to appear in emission, the total flux spectrum ought to be dereddened with $A_V >7$ mag. From the reddening law of \citet{gordon2023} (see Appendix~\ref{ism}), the extinction at 10 \micron\ becomes 0.57~mag and therefore the optical depth $\tau_{\rm 10\mu m}$ is 0.52. This is somewhat higher than what is found from our analysis above, considering that the correlated spectra had already been dereddened for extinction $A_V=2.5$~mag. If the interstellar extinction is not taken into account, our analysis shows that the maximum $\tau_{\rm \lambda}$ is 0.35 on these large spatial scales. 

As mentioned above, we argued that the absorption feature originates from the disk of the HBe star and it is contained within a radius of $<40$~au. This may contradict the scenario of a carved-out dusty cocoon \citep[e.g.,][]{canovas2012}, while the blueshifted jet emission \citep{poetzel1989,stelzer2009,whelan2010,antoniucci2016,liimets2023} should originate from a much smaller region (<10~au). Furthermore, we note the lack of any large-scale, circumbinary warm dust emission in the reconstructed maps.

Evidently, the variability of the silicate feature, as suggested by the MATISSE results, cannot be fully explored with simple analytical models as used in literature studies, but it should be treated with radiative transfer simulations to study the disk's geometry and the dust's chemical composition, surface density, and grain size. We aim to explore this computationally expensive task in future work.

\subsection{Binary orbit}\label{sec:binary}

\begin{figure*}[!btp]
    \centering
    \includegraphics[width=\linewidth]{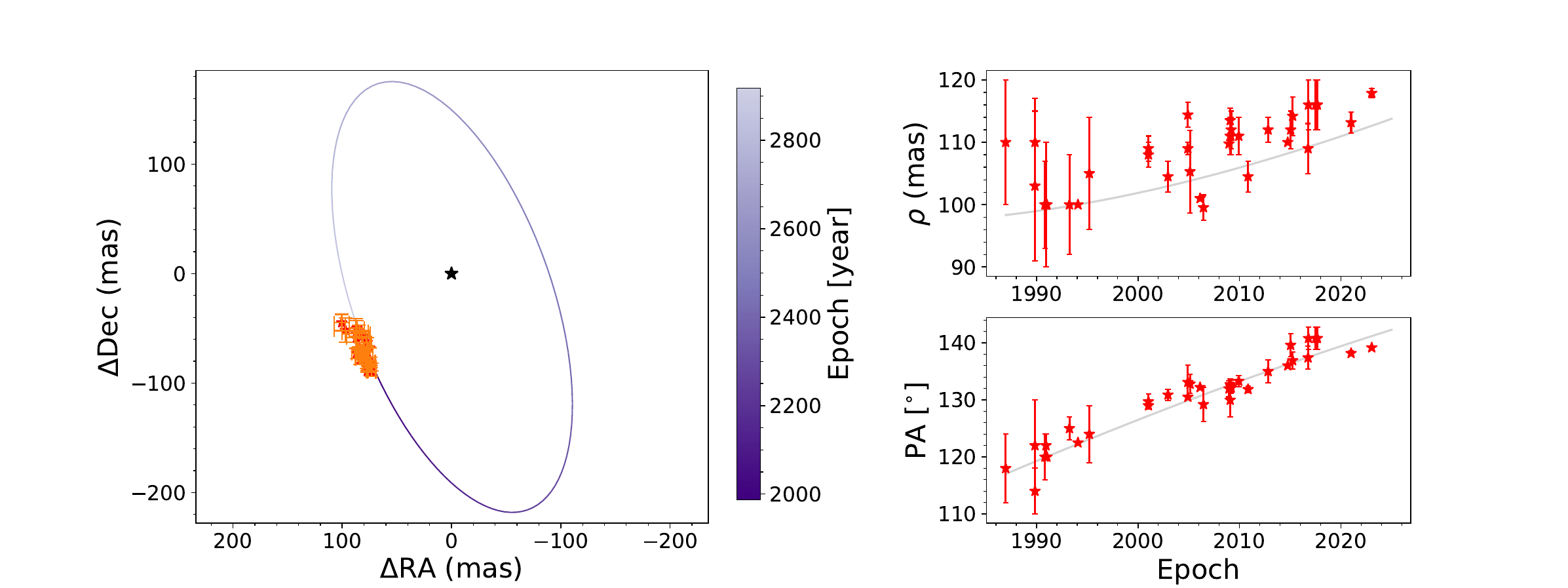}
    \caption{Best-fit orbital solution from {\tt orbitize} using relative astrometry for Z~CMa (Table~\ref{tab:relastro}). {\it Left panel} shows one orbit (purple line, color-coded per year as shown in the color bar) with the relative astrometry of the FUor companion (orange points) and the location of the HBe star (asterisk). {\it Right panels} show the solution of the same orbit (gray lines) against the angular separation (top) and PA (bottom) that is also included in the left panel.}
    \label{fig:myorbit}
\end{figure*}

Using the MATISSE astrometry and archival data (Table~\ref{tab:relastro}), and based on an earlier analysis by \citet{antoniucci2016} and \citet{bonnefoy2017}, we explored the binary's orbit. Figure~\ref{fig:myorbit} shows the relative position of the FUor (orange points) with respect to the HBe star (black star), which is set at the origin (0,0). These relative astrometric measurements are essential in deriving the orbital elements and and estimating a mass ratio of Z~CMa, assuming a fixed mass for one of binary components. It is worth noting that there have been no previous observational attempts to estimate masses for each component; for example, by measuring the radial velocities of each star through high-spectral-resolution studies. Since this is a wide binary, we expect that it would take several decades to collect radial velocity data. Hence, relative astrometry can offer the best solution here.

We used the tool {\tt orbitize}\footnote{\url{https://orbitize.readthedocs.io/}} by \citet{orbitize} to estimate the orbital elements with the Markov chain Monte Carlo (MCMC) solution. We opted for 200 walkers with 20 temperatures and 50,000 steps per walker after allowing a burn-in phase of 10,000 steps. The initial parameters required for this model were the relative astrometry, the source's parallax, and the total mass of the binary. As stated earlier, we adopted the distance estimate of \citet{dong2022}, which we converted to a parallax ($0.85\pm0.05$ mas). This parameter is fixed as a prior to minimize degeneracies. We also tested models where the parallax is a free parameter, and it was found to range within the estimates of \citet{dong2022}. Since the mass ratio of the system is unknown, we opted for a range of $10\pm5$~M$_\odot$ for the total mass, which should include the entire range of possible masses for each star.  Figure~\ref{fig:myorbit} shows the best-fit orbital solution from {\tt orbitize}. The median values of the orbital elements are the semi-major axis, $a$ ($245.6^{+35.7}_{-42.9}$ au), the eccentricity, $e$ ($0.172^{+0.058}_{-0.057}$), the inclination, $i$ ($65.9^{+3.0}_{-4.4}$ deg),  the argument of the periastron of the secondary's orbit, $\omega$ ($142.1^{+9.2}_{-30.6}$ deg), and the longitude of ascending node, $\Omega$ ($200.1^{+1.7}_{-1.8}$ deg). The system's total mass, $M_{\rm total}$, is estimated at $16.4^{+2.1}_{-2.3}$~M$_\odot$. All given errors represent the 68\% confidence intervals from the MCMC fitting (Fig.~\ref{fig:corner}). The orbital period is estimated at $950^{+218}_{-256}$ years. The parameters $\omega$ and $\Omega$ cannot be constrained further in the absence of radial-velocity information \citep[e.g.,][]{nowak2024}.

Our fit provides a conservative estimate of the system's total mass. Assuming that the FUor companion is expected to be $< 2$~M$_{\odot}$, the mass of the primary HBe would range from 12 -- 16~M$_{\odot}$, and thus at the most massive end of its sequence. This could make Z~CMa an early precursor of a supernova explosion.

There exist indirect estimates of the Herbig star's mass; however, several other literature studies have inadvertently used wrong distances\footnote{It is worth mentioning that Gaia DR3 RUWE$>35$ for Z~CMa, while the proper motion of its photocenter is yet to be analyzed. } and/or cumulative photometric measurements in the SED analysis that did not take into account the source's variability \citep{alonsoalbi2015, 2018A&A...620A.128V, marcosarenal2021, 2021A&A...650A.182G, stapper2022, grant2023, 2023AJ....166..183V, thomas2023}. \citet{vdancker2004} argued for a 16~M$_{\odot}$ mass for the Herbig star and approximately 3~M$_{\odot}$ for the FUor. More recently, \citet{fairlamb2015} estimated a mass of $11.0\pm1.7$~M$_\odot$ for the HBe star by fitting theoretical spectra to their XSHOOTER results and for their adopted distance of $1050\pm210$~pc, which is closer to the lower end of our mass estimate for the HBe.

\subsection{Nondetection of a tertiary companion}

The detection of the streamer in NIR polarimetric imaging from the last decade \citep{millangabet2002,canovas2015,liu2016} had been attributed to a stellar fly-by with a probable identification of the intruder at 1.3-mm by \citet{dong2022}. More recently, \citet{zurlo} showed that the streamer feature is in fact part of the extended outflow of the system, while they disproved the presence of said intruder (otherwise a nondetection). The authors identified another background source at approximately 0\farcs 75 northeast of the Z~CMa binary as a tertiary companion. They claimed that the new source, evident as scattered light in the $H$ band, as well as in the 1.3-mm continuum \citep{dong2022}, was marginally detected in the $L'$ band. However, one could argue that the last detection is somewhat similar to spurious noise from PSF subtraction or that the source lies inside the jet of Z~CMa. 

The current relative astrometry does not allow for an evaluation of multiplicity for Z~CMa beyond the known FUor companion. Moreover, this potential tertiary companion falls outside the MATISSE FOV in the $L$ band, while in image reconstructions with the broader FOV of the $N$ band, no other source is identified fainter than the FUor companion, the latter being more than ten times fainter than the primary Herbig star in the $N$ band.

\section{Conclusions}

We present our MIR interferometric imaging results from MATISSE/VLTI of the young eruptive stars binary Z~CMa, which were obtained during the 2023 EXor-like major outburst of the Herbig primary. This is the first part in a series of papers from these MATISSE observations.

We provide mono-chromatic images at several wavelengths essentially corresponding to continuum emission in the $L$, $M$, and $N$ bands. The binary is resolved in all bands, and extended circumstellar emission is evident around each star. From the images we calculated relative brightness ratios in all three bands and find that the FUor is much fainter than the Herbig star, and more than 10x fainter in the $N$ band in particular. 

Optical depth maps indicate a nonuniform distribution of silicate dust on large spatial scales, or else within a region of 40 -- 130~au in diameter, where the optical depth increases monotonically toward smaller radii. This distribution is more probably associated with the bright Herbig star and its protoplanetary disk and should be explored further with dust radiative transfer simulations. The dusty cocoon surrounding the Herbig star that was alleged by earlier NIR polarimetric imaging studies may be transparent in the MIR since emission is clearly seen from the protoplanetary disks of both binary components. We do not find any evidence for substructures, such as gaps, in the MATISSE images of the HBe disk, but this could be due to the limited angular resolution taking into account the object's distance.

The MIR-emitting region of the FUor companion's disk is rather small in all spectral bands compared to that of the primary, at an approximately constant angular size of $\le 15$~mas, which translates to a physical radius of $< 9$~au. This is congruent with recent MATISSE results on FUor disks, which allude that these are rather compact, especially when the system is returning to quiescence. The HBe disk extends to radii of $\le 8$~au in the $L$ band and $\ll 35$~au in the $N$ band. At longer wavelengths and lower flux scales, the disk is marginally smaller than the 1.3-mm optically thin emission in ALMA studies \citep{dong2022}. The disk position angles at 152.5\degr$\pm\,$12.7\degr\ and 156.5\degr$\pm\,$12.1\degr\ and inclinations at 33.5\degr$\pm\,$5.7\degr\ and 26.6\degr$\pm\,$3.8\degr\, for the FUor and HBe, respectively, suggest that the disks are aligned, and they agree with the orientations extrapolated from the jets \citep{whelan2010,antoniucci2016}.

From the MATISSE reconstructed images, we calculated a binary separation of $117.88\pm0.73$ mas at a position angle of 139.16\degr\,$\pm$ 0.29\degr . This is complementary to approximately 20 years of relative astrometry data. Our orbital fitting provided estimates for the total mass of the system $M_{\rm total}=16.4^{+2.1}_{-2.3}$~M$_\odot$ and a period approximately 950 years. The fit orbital inclination ($i\sim66^{\rm o}$) is somewhat different to the calculated inclinations, within the large uncertainties, of the disks from 1.3-mm and 9-mm images \citep{dong2022}, as well as those estimated from the MATISSE images. Due to the large uncertainties, we cannot ascertain whether there is a misalignment between the millimeter continuum and MIR wavelengths. We expect that the orbital solutions can be constrained better with future astrometric observations (e.g., Gaia DR4).

Z~CMa is a highly complex system with an evolving Herbig star, and an FUor companion that may have returned to quiescence. Future MIR observations of the system during the Herbig star's quiescence with MATISSE/VLTI and METIS/ELT may reveal more regarding extended circumbinary emission. We encourage new ALMA observations at higher angular resolution to investigate the discrepancy between MATISSE and ALMA disk sizes. In future works as part of this series, we intend to employ radiative transfer methods to investigate the geometric structure and chemical composition of the HBe disk, and exploit the spectro-interferometric MATISSE data.

\begin{acknowledgements}
We would like to thank the anonymous referee for their feedback that helped improve this manuscript. This work received funding from the Hungarian NKFIH OTKA project no. K-132406 and K-147380, and the NKFIH Excellence grant TKP2021-NKTA-64. This work was also supported by the NKFIH NKKP grant ADVANCED 149943. Project no. 149943 has been implemented with the support provided by the Ministry of Culture and Innovation of Hungary from the National Research, Development and Innovation Fund, financed under the NKKP ADVANCED funding scheme. FCSM received financial support from the European Research Council (ERC) under the European Union’s Horizon 2020 research and innovation programme (ERC Starting Grant “Chemtrip”, grant agreement No 949278).

Based on observations collected at the European Organisation for Astronomical Research in the Southern Hemisphere under ESO programmes: 0106.C-0501(B), 0108.C-0385(D), and 0110.C-4209(A). Spectroscopic observations were obtained as Visiting Astronomer at the Infrared Telescope Facility, which is operated by the University of Hawaii under contract 80HQTR19D0030 with the National Aeronautics and Space Administration

This research has made use of the Jean-Marie Mariotti Center \texttt{OImaging} service \footnote{Available at http://www.jmmc.fr/oimaging} part of the European Commission's FP7 Capacities programme (Grant Agreement Number 312430), and \texttt{LITpro} service co-developed by CRAL, IPAG and LAGRANGE\footnote{LITpro software available at http://www.jmmc.fr/litpro}. We acknowledge with thanks the variable star observations from the AAVSO International Database contributed by observers worldwide and used in this research. Based on data from the OMC Archive at CAB (INTA-CSIC), pre-processed by ISDC and further processed by the OMC Team at CAB. The OMC Archive is part of the Spanish Virtual Observatory project. Both are funded by MCIN/AEI/10.13039/501100011033 through grants PID2020-112949GB-I00 and PID2019-107061GB-C61, respectively.
\end{acknowledgements}
\bibliographystyle{aa}
\bibliography{ZCMA_LIBS}

\begin{appendix}

\section{Interstellar extinction}\label{ism}

The current literature includes several estimates for the interstellar extinction, $A_V$, toward Z~CMa (Table~\ref{tab:ext}). The analysis of Bayestar19 \citep{green2019} and STILISM \citep{2019A&A...625A.135L} indicates the presence of a foreground cloud within 1~kpc. This could explain the extremely high extinction values estimated by IRAS and Planck satellites, whose wide beams would have engulfed not only the foreground cloud but also dust from the surrounding CMa R1 star forming region. The large uncertainties by \citet{connelley2018} are a result of their fitting the spectrum of Z~CMa to that of FU~Orionis. The protostars themselves are dusty, therefore we cannot exclude that the remaining estimates in Table~\ref{tab:ext} may be affected by circumstellar and/or circumbinary extinction as well. However, the contribution for each protoplanetary disk is unclear, since both are visible in the infrared and (sub)millimeter wavelengths. The large uncertainties by \citet{connelley2018} are a result of their fitting the spectrum of Z~CMa to that of FU~Orionis, while \citet{hinkley2013} attempted to fit the NIR photometry of each star with reddened blackbody spectra to estimate circumstellar extinction. 

In this work, we adopt an interstellar extinction of $A_V = 2.5$ mag. This is the average of all values where $A_V \leq 5$ mag in Table~\ref{tab:ext}, and it is in agreement with the extinction used in \citet{sag2020}. When necessary, we used the extinction law of \citet{gordon2023} to deredden spectrophotometry with $R_V=3.1$, which for our adopted $A_V$, results in extinction of $A_{\rm 10\mu m}/A_V \simeq 0.08$ or else $A_{\rm 10\mu m} \simeq 0.2$ mag for the silicate feature.

\begin{table}[hbtp]
    \centering
        \caption{Interstellar extinction around Z~CMa.}
    \begin{tabular}{lcl}
    \hline\hline
        Source & $A_V$ (mag) & Notes\\
        \hline
        \citet{2011ApJ...737..103S} & 10.3 & IRAS \\
        GALEXtin, \url{www.galextin.org} & 18.3 & Planck \\
        \citet{hinkley2013} & 5.2-10 & fitted \\
        \citet{connelley2018} & 7.1$\pm$8.0 & FU~Ori \\
        \citet{hi4pi} & 3.0 & HI 21cm \\
        {\citet{2019A&A...625A.135L}} & 1.2 & STILISMv.1 \\
        UKSSDC & 4.6 & Galactic $N_H$ \\
        \citet{green2019} & 2.0 & Bayestar19 \\
        \citet{sag2020} & $\sim2.5$ & \\
        \citet{stelzer2009} & 2.4-4.6 & literature\\
        \citet{carvalho2022} & 0.96 & DIBs \\
        \citet{fairlamb2015} & 3.37 & fitted \\
        \citet{herbst1978} & 0.68 & foreground \\
        \hline
    \end{tabular}
    \label{tab:ext}
\end{table}


\section{MATISSE data quality}\label{sec:matqual}

\begin{figure*}
    \centering
    \includegraphics[width=0.9\linewidth]{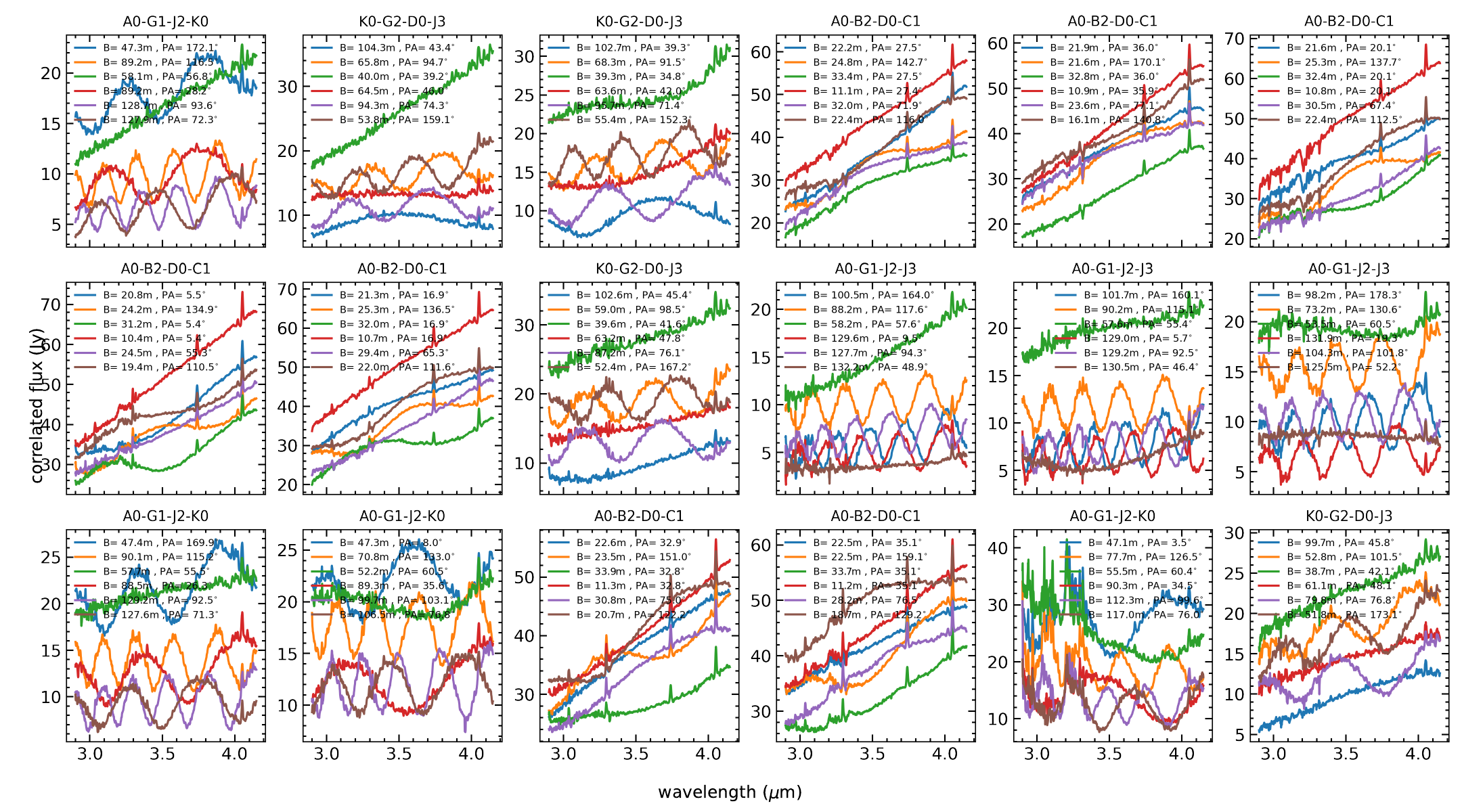}
    \caption{MATISSE $L$-band correlated fluxes. These are color-coded w.r.t. baseline (length, PA) as listed within each panel. From left to right and top to bottom each panel corresponds to sequential observations with the array configurations listed in Table~\ref{tab:matlog}, that is ids=3, 5-13, 15-16, and 18-23. }
    \label{fig:fcorrL}
\end{figure*}

\begin{figure*}
    \centering
    \includegraphics[width=0.9\linewidth]{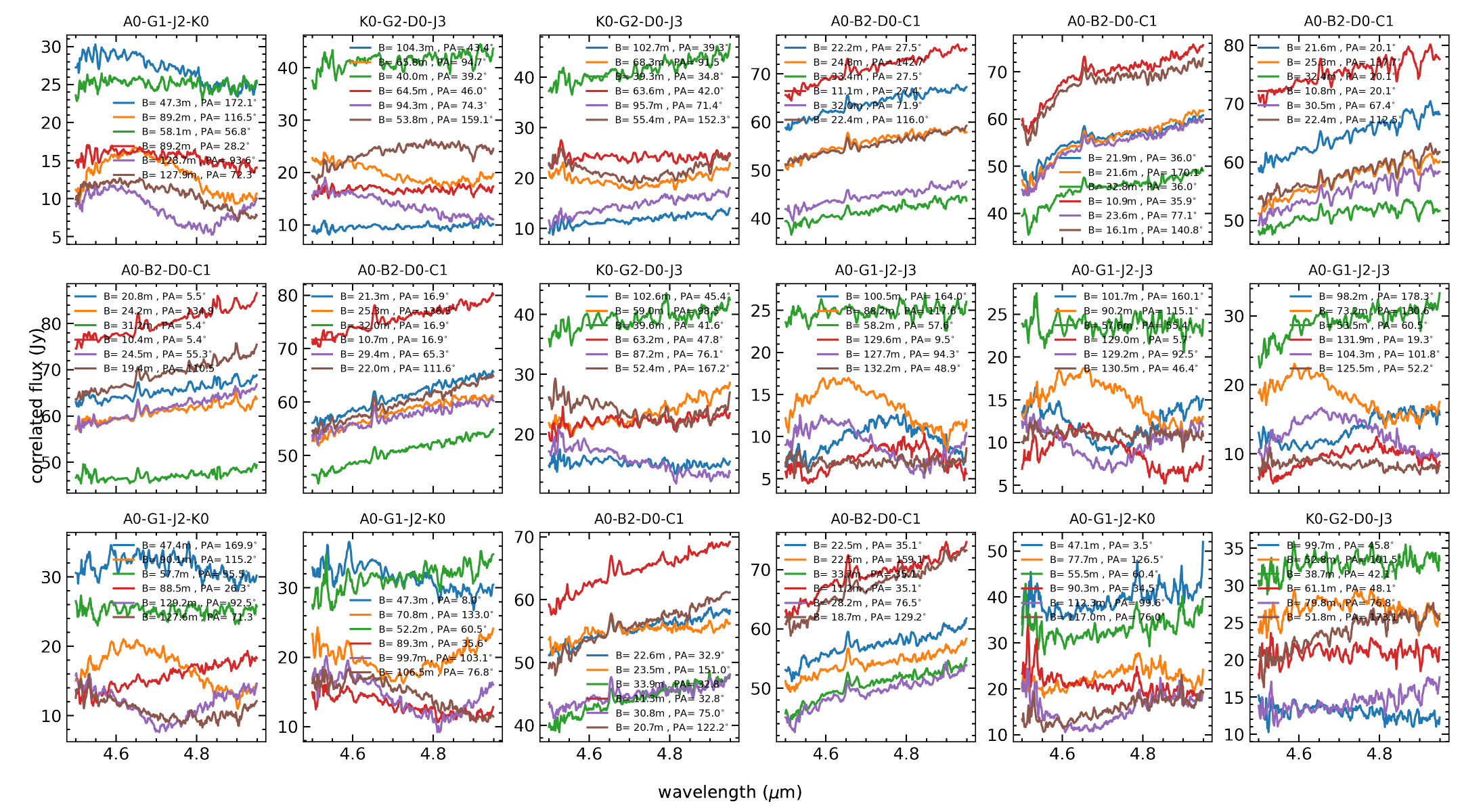}
    \caption{As in Fig.~\ref{fig:fcorrL} for the $M$-band correlated fluxes.}
    \label{fig:fcorrM}
\end{figure*}

A summary of the MATISSE/VLTI data is shown in  Table~\ref{tab:matlog}. Observational blocks that were not successful and/or aborted due to adverse atmospheric conditions and/or technical faults at the telescope are also indicated as `X', and these datasets are discarded from this work. The GTO observations from 2021 were obtained in MED-L, LOW-L, and HIGH-N (id=1) and LOW-L and LOW-N (id=2) spectral resolutions, while the imaging observations were all obtained in MED-L and LOW-N (ids= 3 - 23). Only the LOW-L and LOW-N data from id=1 were used in this work.
\paragraph{GRA4MAT and $LM$ band.} In the majority of cases, chopped photometry failed in the $LM$ band even during the best observing conditions (i.e., atmospheric coherence times $t_{\rm coh}\geq 3$~ms) and even with the bright calibrator ($\alpha$ CMa). It is unclear why this is the case for GRA4MAT even with a bright science target like Z~CMa ($F_{\rm3.5\mu m}=40$ Jy). Therefore, only the non-chopped $LM$-band (and subsequently $M$) data is used in this work. This issue does not affect the $N$-band data. To validate the photometric accuracy of the non-chopped data from the 1.8-m ATs in all bands, we compared the averaged MATISSE spectra against the archival $M$ and $N$ band photometry from the Midsource Space Experiment (MSX) and AKARI space telescopes, as well as our own SpeX/IRTF spectroscopy. We found that all are in good agreement (Fig.~\ref{fig:specsed}). We also note that Z~CMa saturates (NEO)WISE photometry and therefore these measurements are unreliable.
\paragraph{GRA4MAT and $N$ band.} The standard MATISSE pipeline calibration (ver.1.7.5 and later) failed for the case of the $N$-band data with GRA4MAT. We suspect that this was caused by the optical path difference (OPD) selected by GRAVITY for GRA4MAT (co-phasing) in the $N$-band. To correct for this, we reduced the $N$-band data with an earlier version of the pipeline (ver. 1.6) that is used in the standalone MATISSE mode. This issue does not affect the $LM$-band data. Furthermore, GRA4MAT does not record simultaneous photometry for the $N$ band. This can be recovered instead from the interferometric data by dividing the correlated fluxes with the visibilities. The quality of this data has been verified to be in good agreement against archival MIDI spectra \citep{varga2018} and archival photometry (Fig.~\ref{fig:specsed}).
\paragraph{Rotational flip in the closure phases.} The phase sign ambiguity for the $N$-band data that was mentioned by \citet{gamezrosas2022} was not resolved in the MATISSE DRS pipelines used for this work (ver. 1.7.5 and earlier). This was confirmed in our data, since the orientation of the Z~CMa binary is well known (see Sect.~\ref{sec:orb}). Therefore, all our models and image reconstruction results in the $N$ band have been rotated by 180\degr\ to correct for this issue.\footnote{ The latest version of the pipeline (ver.2.0) has corrected this issue, while the remaining features of the pipeline are the same. But since there is a simple solution of rotating our images by 180\degr, we refrained from repeating the data reduction with the new pipeline.}
\paragraph{Choosing calibrators.} Sirius ($\alpha$ CMa) was selected as a calibrator for our small array observations (ids: 7-11, 20, and 21 in Table~\ref{tab:matlog}) due to its high relative brightness for the ATs. Sirius is a binary system, but a wide one with a current separation $\gg 2$" and therefore outside the ATs FOV. The white dwarf companion (Sirius B) is not known to emit in the MIR. Unfortunately, due to its proximity to the Sun, the A-type main sequence star (Sirius A) can be resolved with the small array \citep[$\diameter\approx6.14$ mas;][]{jsdc}. Moreover, the hydrogen spectral signatures of Sirius are clearly detected in the calibrated data (single-dish spectra and correlated fluxes). Although the hydrogen lines in Sirius A's spectrum are in absorption, since the calibrated data are processed by dividing the response of the science target with that of the calibrator, they appear here as emission lines (cf. Fig.~\ref{fig:fcorrL} and \ref{fig:fcorrM}). Hence, when necessary in our analysis, we either subtract the hydrogen lines from the small array data by replacing them with the average continuum or we select a continuum region that does not include these lines. Image reconstruction is not affected, since this is based only on the closure phases and visibilities where we do not see any influence by Sirius' spectral response. %

\begin{table*}[bhtp]
\caption{Summary for MATISSE/VLTI observations. }\label{tab:matlog}
\centering
\begin{tabular}{ccccccccc} 
\hline\hline             
id & Date and UT & Array & Stations & Seeing (\arcsec) & $\tau_0$ (ms) & Phase & Calibrator & Comments \\
\hline
1 & 2021-01-23T04:50    & medium & K0-G2-D0-J3  & 0.93 & 3.75 & Q &     N: HD48915 & GTO \\
\multicolumn{7}{c}{} & L: HD58972 & \\
2 & 2021-11-25T08:32    & medium & K0-G2-D0-J3  & 0.54 & 8.17 & Q &     HD39853 & GTO \\ \hline
3 & 2022-11-29T07:20    & large  & A0-G1-J2-K0  & 1.04 & 3.21 & O &     HD48217 & C \\
4 & 2022-12-25T05:46    & medium & K0-G2-D0-J3  & 0.80 & 3.07 & O &     HD48217 & X \\
5 & 2022-12-27T06:18    & medium & K0-G2-D0-J3  & 1.51 & 1.90 & O &     HD48217 & B \\
6 & 2022-12-28T05:11    & medium & K0-G2-D0-J3  & 0.42 & 5.57 & O &     HD48217 & B \\
7 & 2023-01-15T04:10    & small  & A0-B2-D0-C1  & 1.14 & 3.33 & O &     HD48915 & B \\
8 & 2023-01-15T07:17    & small  & A0-B2-D0-C1  & 1.42 & 2.77 & O &     HD48915 & B \\
9 & 2023-01-17T02:54    & small  & A0-B2-D0-C1  & 0.83 & 4.52 & O &     HD48915 & A \\
10 & 2023-01-18T01:06   & small  & A0-B2-D0-C1  & 0.66 & 9.36 & O &     HD48915 & A \\
11 & 2023-01-18T02:29   & small  & A0-B2-D0-C1  & 0.39 & 13.62 & O & HD48915    & A \\
12 & 2023-01-19T05:47   & medium & K0-G2-D0-J3  & 0.62 & 11.01 & O & HD48217    & B \\
13 & 2023-01-25T03:53   & large  &  A0-G1-J2-J3 & 0.71 & 9.16 & O &     HD48217 & A \\
14 & 2023-01-25T05:07   & large  &  A0-G1-J2-J3 & 0.40 & 14.98 & O & HD48217    & X  \\
15 & 2023-01-26T03:12   & large  &  A0-G1-J2-J3 & 0.92 & 5.83 & O &     HD48217 & A \\
16 & 2023-01-26T05:43   & large  &  A0-G1-J2-J3 & 0.73 & 5.90 & O &     HD48217 & A \\
17 & 2023-01-28T01:55   & large  &  A0-G1-J2-J3 & 0.83 & 6.14 & O &     HD48217 & X \\
18 & 2023-02-15T01:56   & large  &  A0-G1-J2-K0 & 1.01 & 4.20 & O &     HD48217 & A \\
19 & 2023-02-15T04:38   & large  & A0-G1-J2-K0  & 0.60 & 9.21 & O &     HD48217 & A \\
20 & 2023-02-25T02:43   & small  & A0-B2-D0-C1  & 0.81 & 4.15 & O &     HD48915 & A \\
21 & 2023-02-25T03:36   & small  & A0-B2-D0-C1  & 0.83 & 4.45 & O &     HD48915 & A \\
22 & 2023-03-20T01:46 & large  & A0-G1-J2-K0    & 0.96 & 4.69 & O &     HD48217 & B \\
23 & 2023-03-25T02:09 & medium & K0-G2-D0-J3    & 1.07 & 3.23 & O &     HD48217 & A \\
\hline
\end{tabular}
\tablefoot{Columns 5 and 6 give the DIMM seeing and the atmospheric coherence time, respectively, at the start of each observation. Column 7 states the variability phase of Z~CMa, (Q)uiescence or (O)utburst. GTO: Guaranteed Time Observations. Data quality is ranked as A (best), B (pass), C (bad), and X (reject). Calibrators (spectral type; angular size): $\alpha$~CMa (A1V; 6.09 mas), HD58972 (K3III; 3.39 mas), HD39853 (K4III; 2.23 mas), and HD48217 (K5III; 2.60 mas). }
\end{table*}

\begin{figure*}
    \centering
    \includegraphics[width=0.9\linewidth]{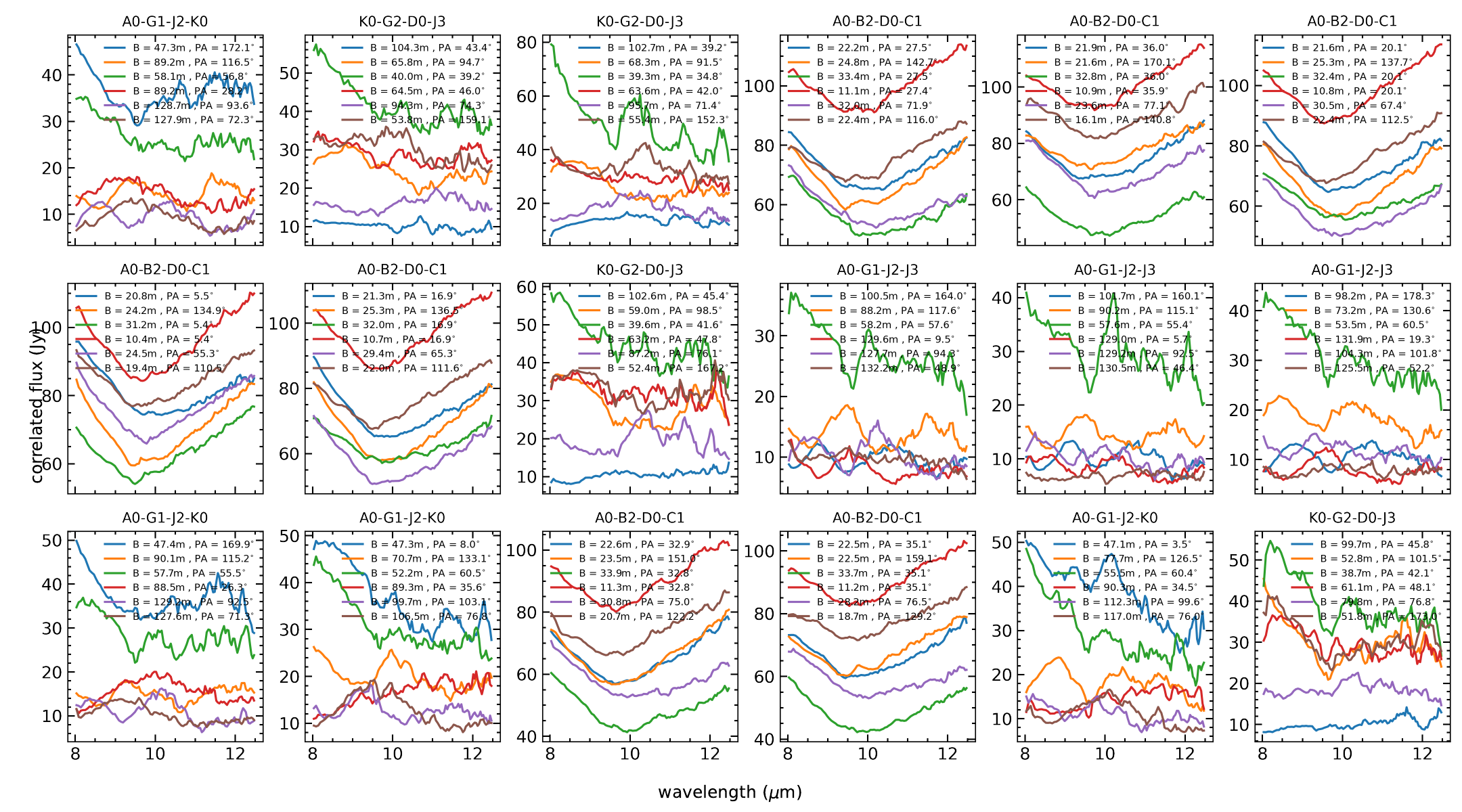}
    \caption{As in Fig.~\ref{fig:fcorrL} for the $N$-band correlated fluxes.}
    \label{fig:fcorrN}
\end{figure*}

\begin{figure*}
    \centering
    \includegraphics[width=0.9\linewidth]{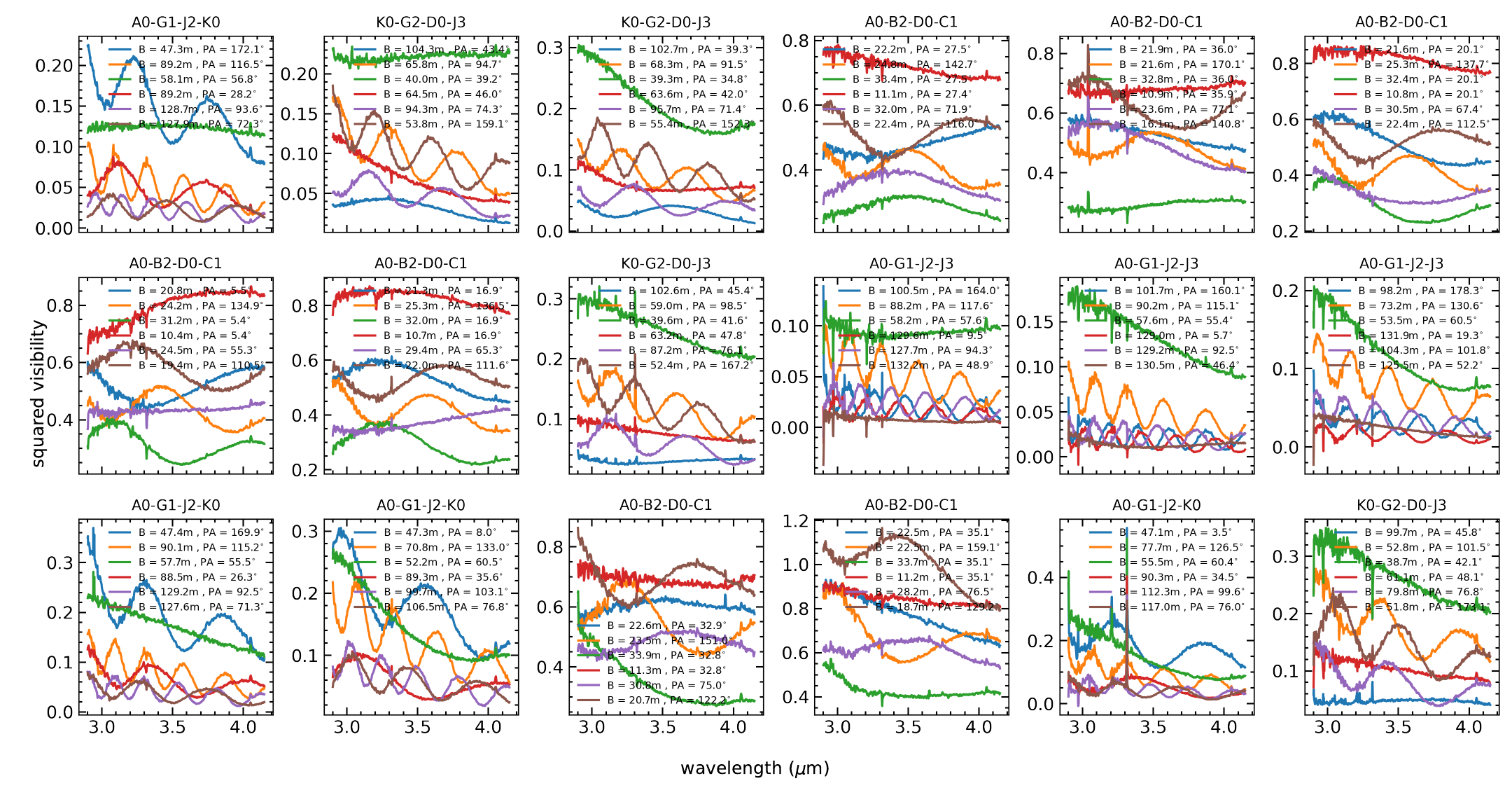}
    \caption{MATISSE $L$-band squared visibilities from the imaging data. The color-coding and sequential order are as in Fig.~\ref{fig:fcorrL}.}
    \label{fig:vis2L}
\end{figure*}

\begin{figure*}
    \centering
    \includegraphics[width=0.9\linewidth]{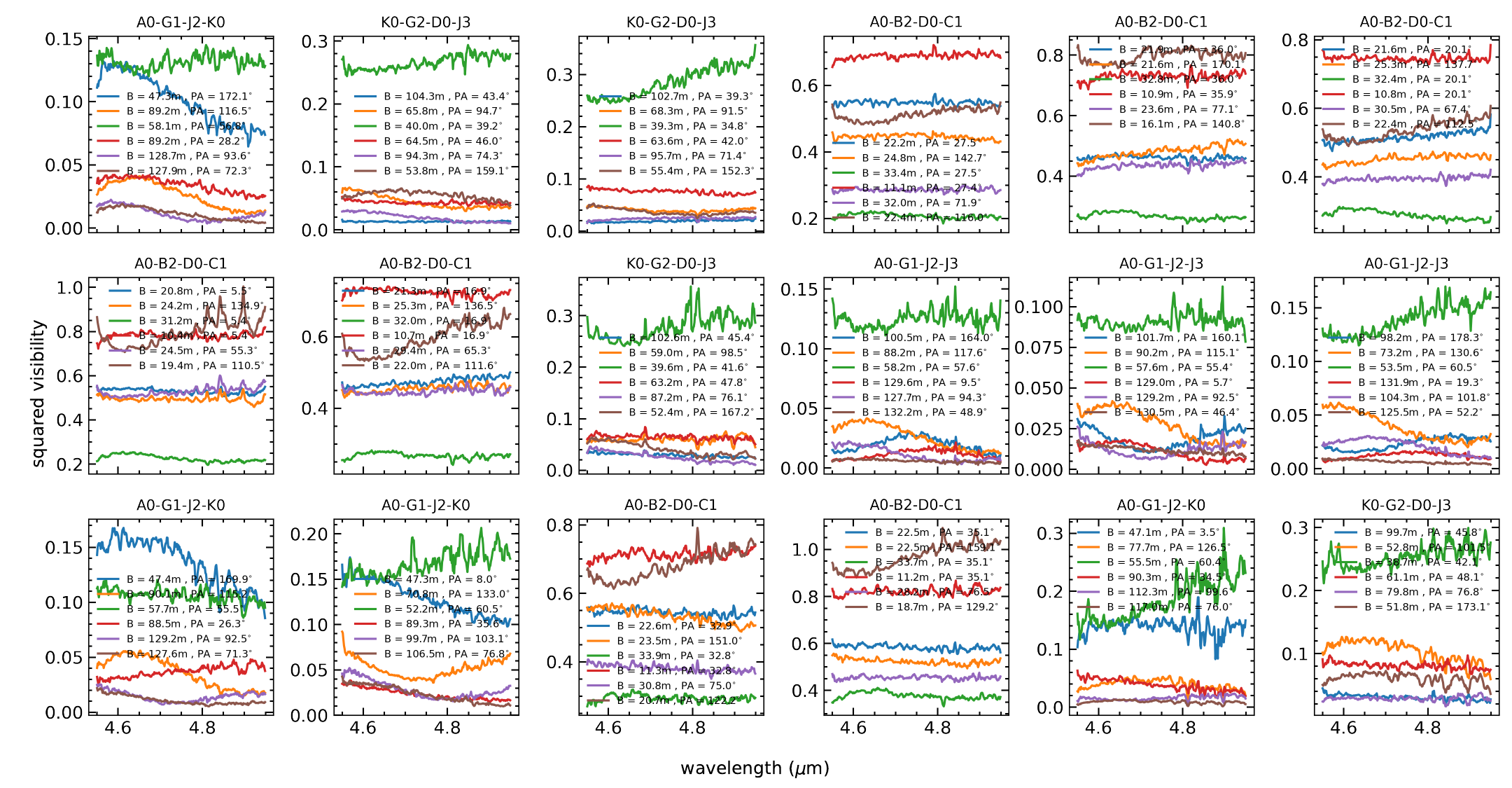}
    \caption{As in Fig.~\ref{fig:vis2L} for the $M$-band squared visibilities.}
    \label{fig:vis2M}
\end{figure*}

\begin{figure*}
    \centering
    \includegraphics[width=0.9\linewidth]{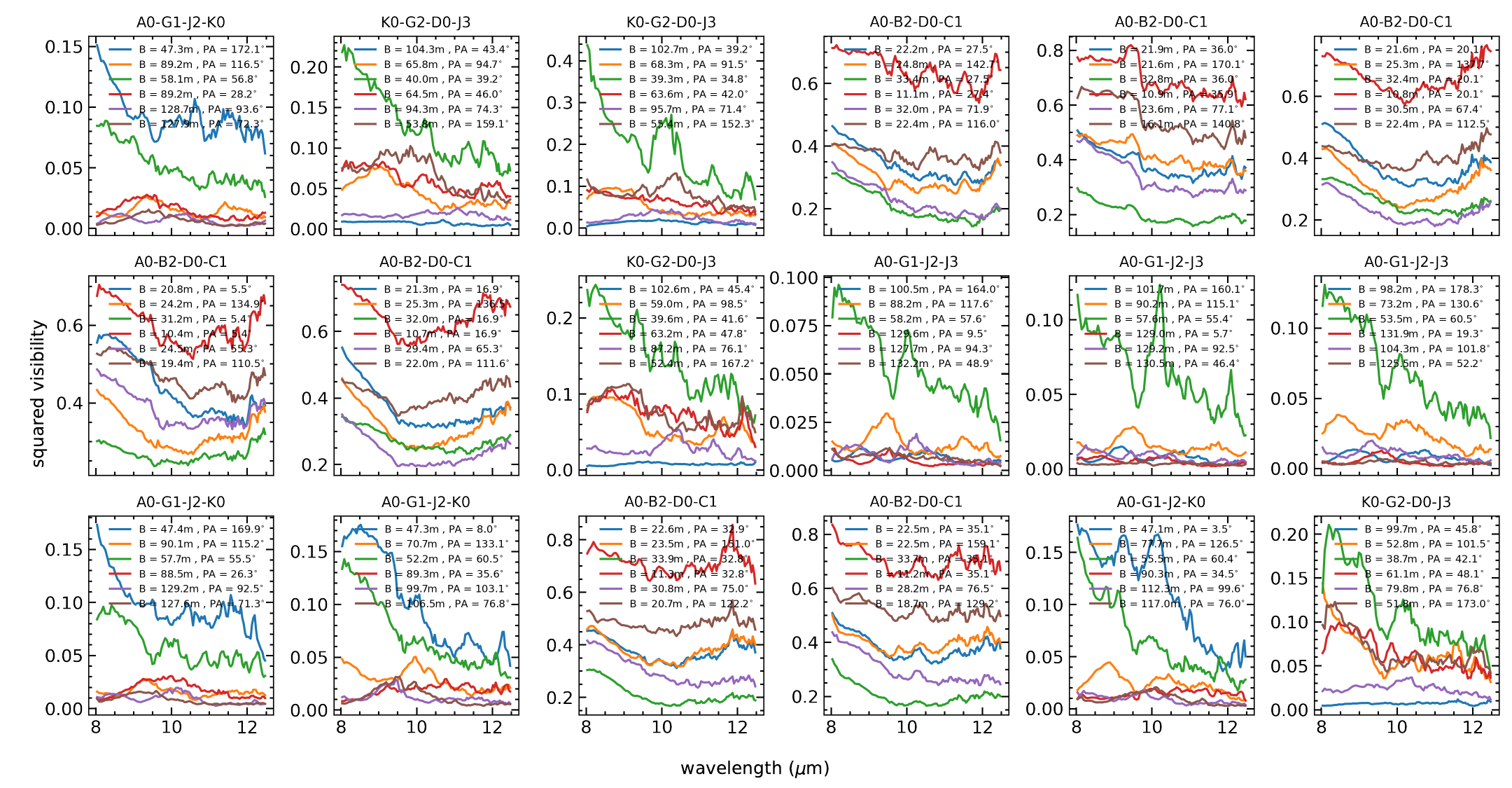}
    \caption{As in Fig.~\ref{fig:vis2L} for the $N$-band squared visibilities.}
    \label{fig:vis2N}
\end{figure*}

\begin{figure*}
    \centering
    \includegraphics[width=0.9\linewidth]{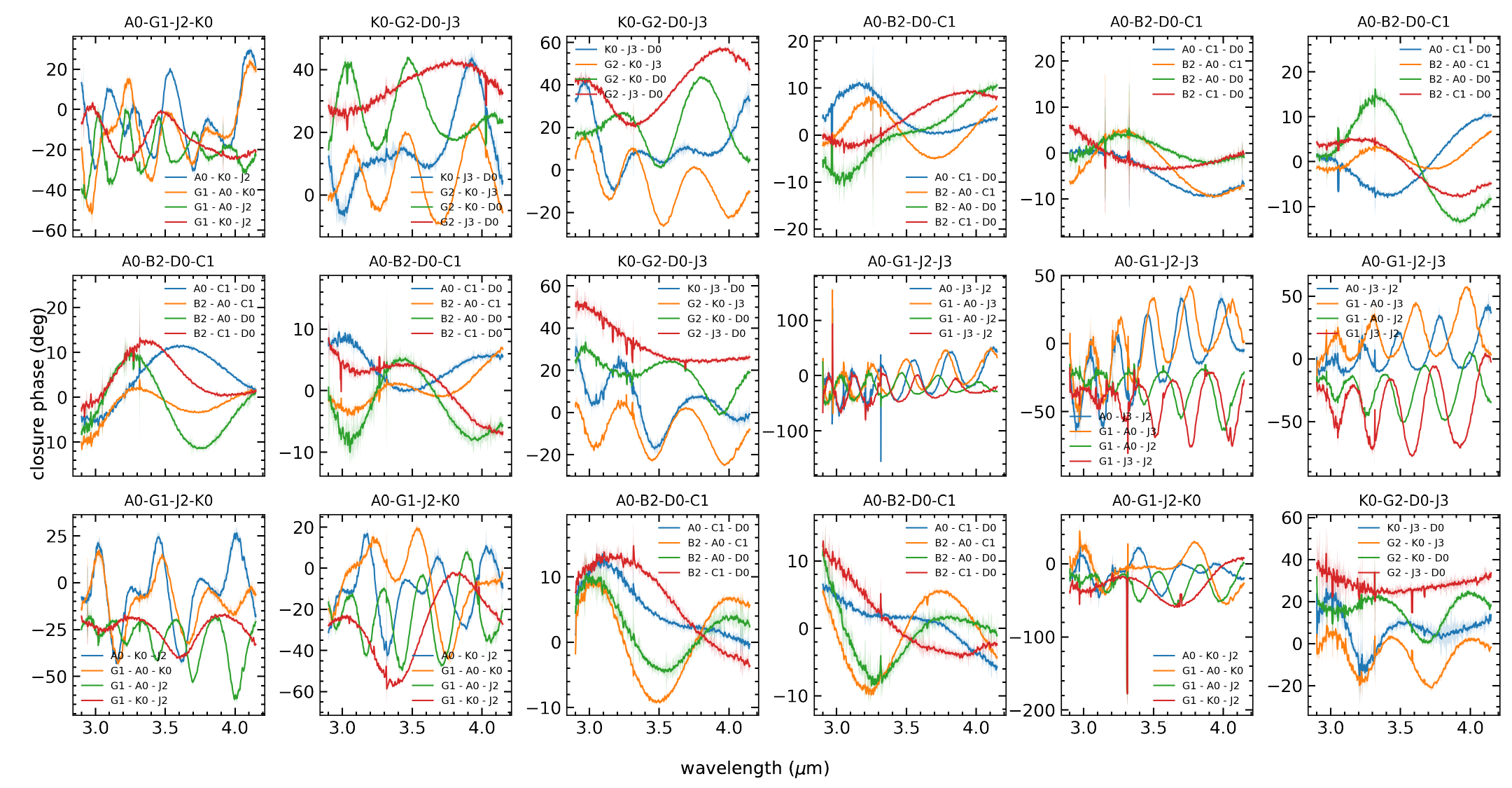}
    \caption{MATISSE $L$-band closure phases from the imaging data. These are color-coded per baseline triangle as shown in each panel. Each panel is sequentially ordered as in Fig.~\ref{fig:fcorrL}..}
    \label{fig:cpL}
\end{figure*}

\begin{figure*}
    \centering
\includegraphics[width=0.9\linewidth]{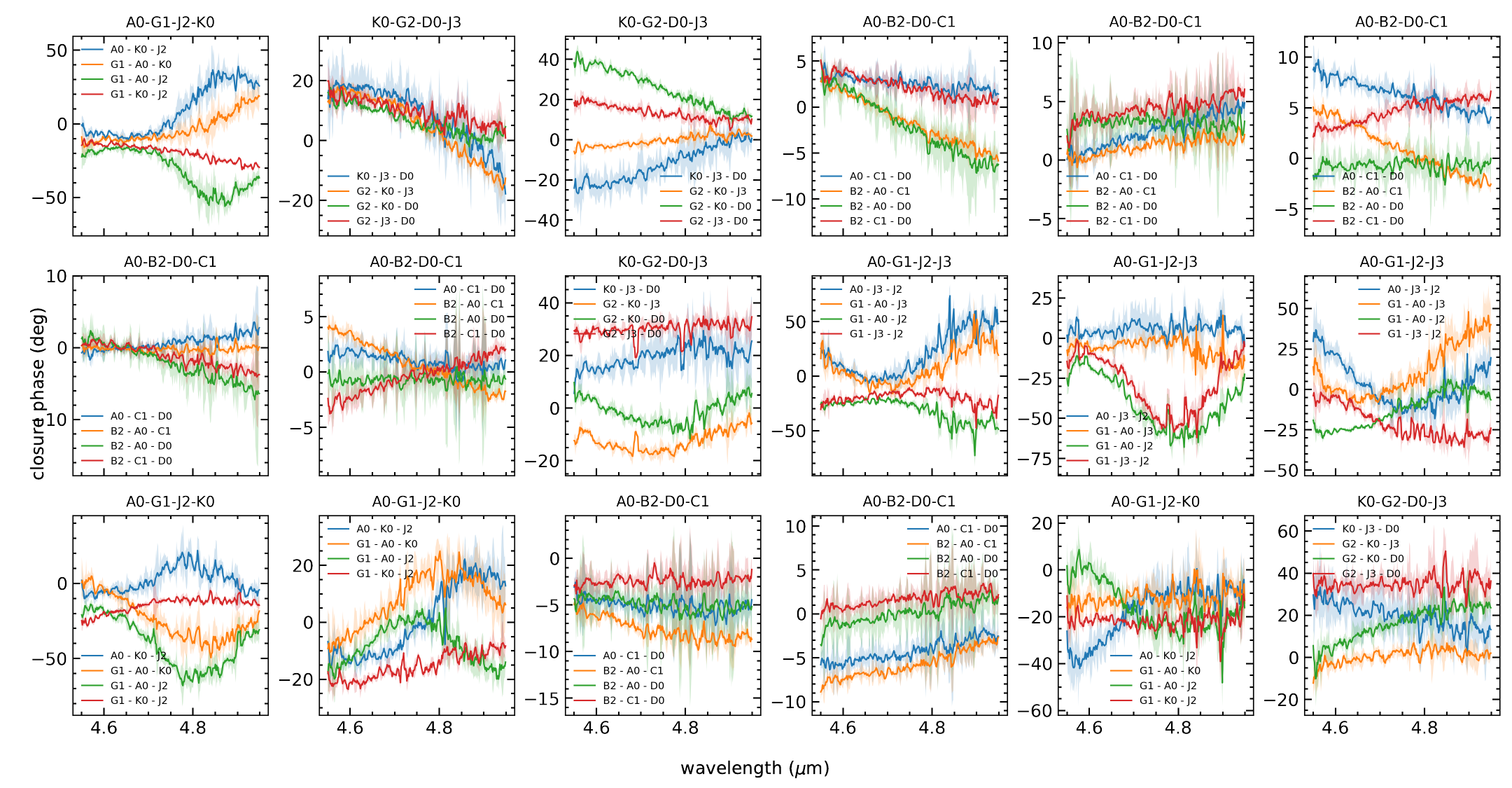}
    \caption{As in Fig.~\ref{fig:cpL} for the $M$-band closure phases.}
    \label{fig:cpM}
\end{figure*}

\begin{figure*}
    \centering
\includegraphics[width=0.9\linewidth]{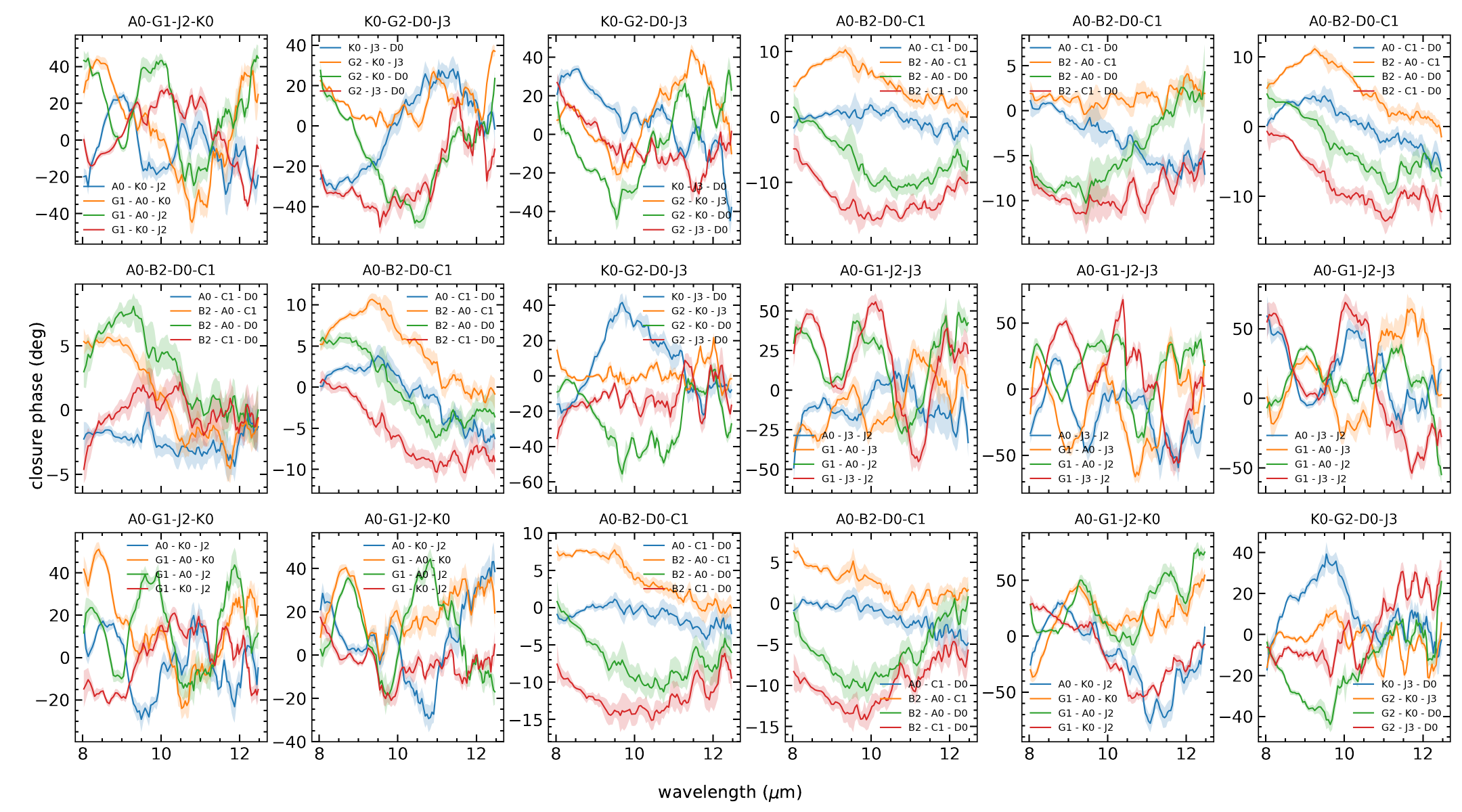}
    \caption{As in Fig.~\ref{fig:cpL} for the $N$-band closure phases. Here, the phase convention is flipped w.r.t. to the $LM$ data.}
    \label{fig:cpN}
\end{figure*}


\section{Image reconstruction methods}\label{sec:imreconmethods}

Unlike radio interferometry where there exist robust image reconstruction algorithms, which take advantage on the well-calibrated information of the visibilities' phase, the sparse $uv$ coverage and absence of absolute (visibility) phase measurements is a well-known problem that affects image reconstruction in infrared interferometry. Furthermore, there exist multiple image reconstruction techniques to tackle such data. It is thus quite common practice to compare images produced from at least two image reconstruction algorithms. Here, we showcase the differences found between BSMEM ver.2.3 \citep{bsmem} and MiRA ver.2.3.2 \citep{mira} algorithms on OImaging for the same set of MATISSE data, which will explain our final choice on selecting an algorithm throughout this work (i.e., MiRA). The comparison is made for the $L$ (3.5\micron) and $N$ (8.5\micron) band data. 

\begin{figure*}[b]
\centering
\includegraphics[width=0.6\linewidth]{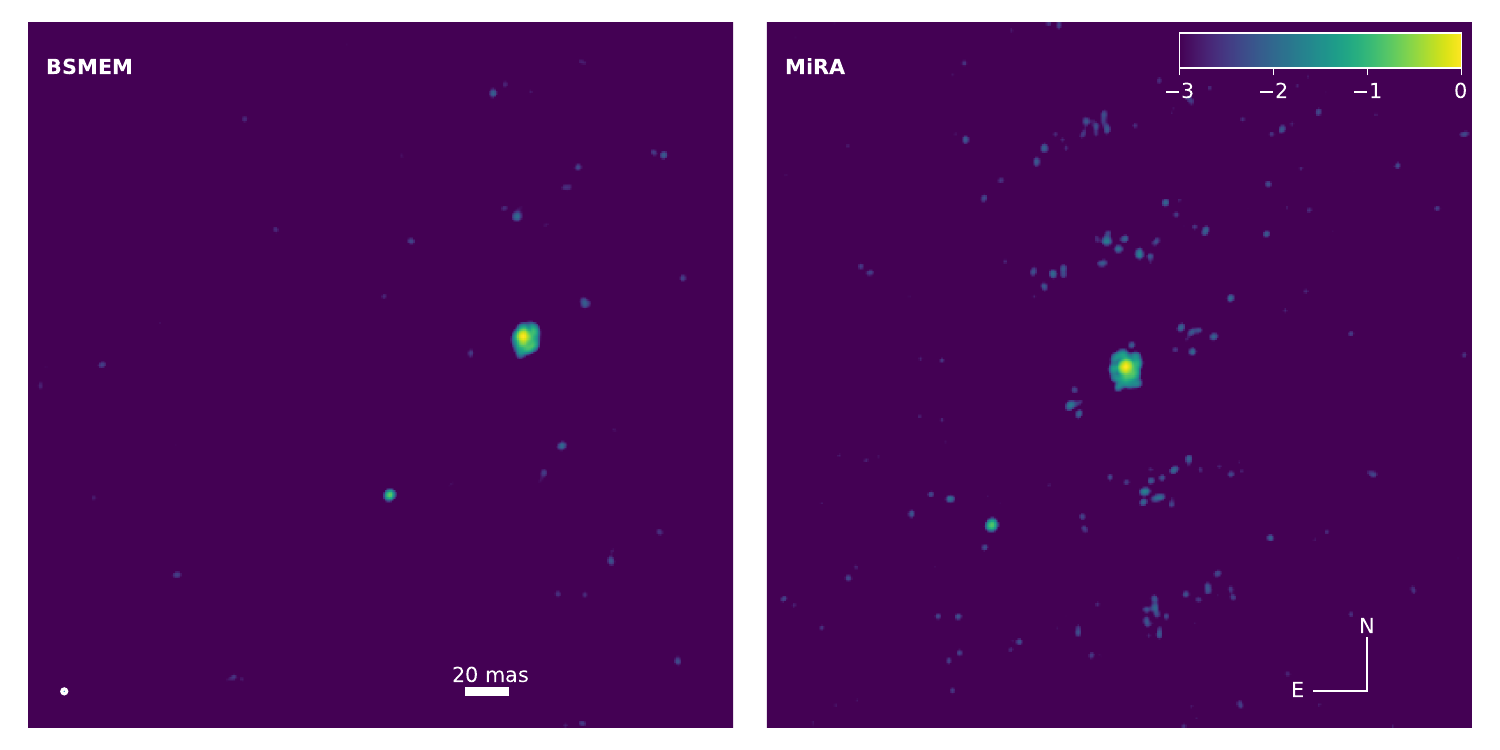}
\includegraphics[width=0.6\linewidth]{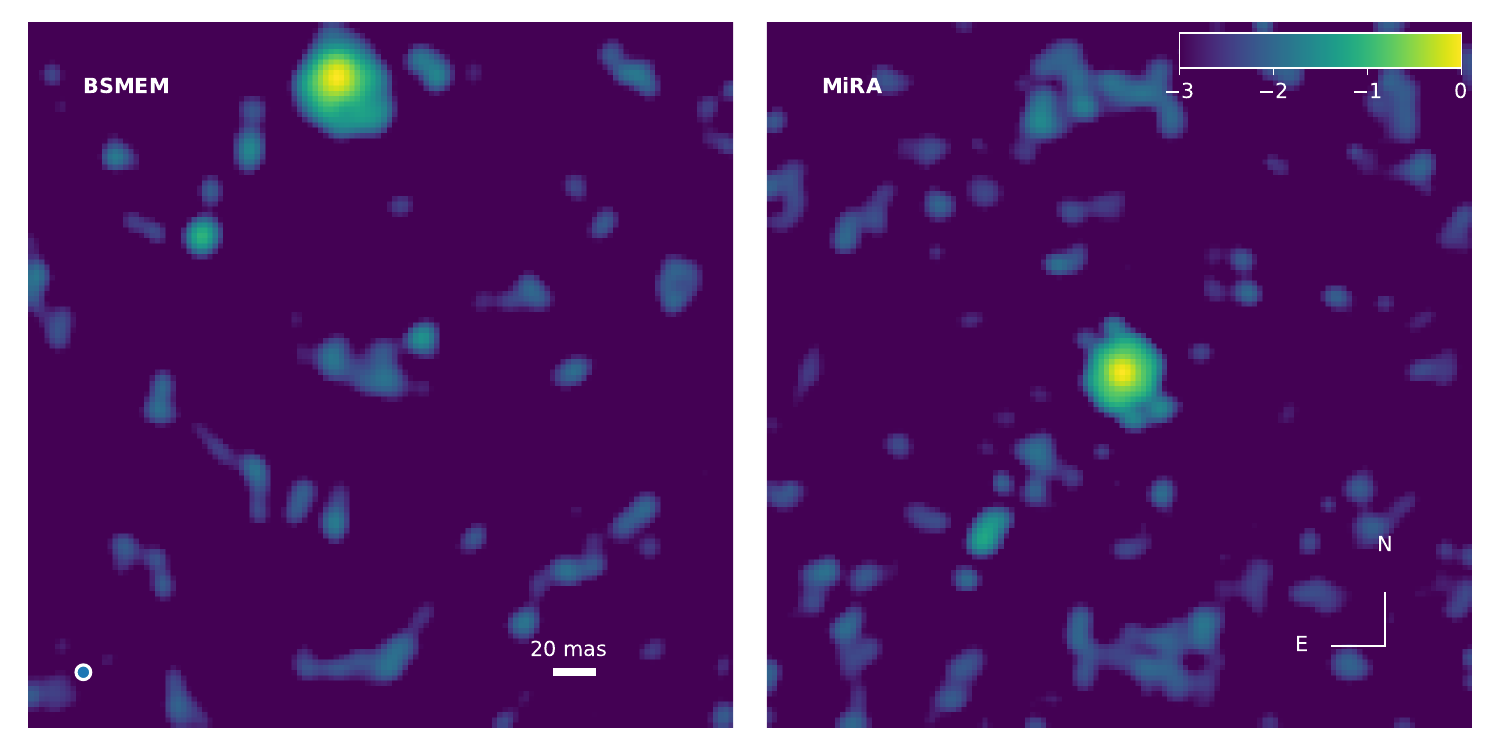}
    \caption{Comparison between BSMEM (left) and MiRA (right) reconstructions at 3.5\micron\ (top row) and 8.5\micron\ (bottom row). Each reconstruction image is normalized to the peak intensity, and scaled in a logarithmic stretch to a minimum level of $10^{-3}$ as indicated in the colorbar. The FOV is  $400\times 400$ mas with north up and east toward left. The theoretical MATISSE beam is shown as a white circle in the lower-left corner of the left panels. A Gaussian smoothing with a FWHM equal to the beam has been applied throughout to enhance the FUor's disk.}
    \label{fig:compareimrecon}
\end{figure*}

Figure~\ref{fig:compareimrecon} shows image reconstructions with BSMEM (left panels) and MiRA (right panels) for continuum emission in the $L$ band (top row; 3.4 -- 3.6 \micron; namely 3.5\micron) and the $N$ band (bottom row; 8 -- 9 \micron; namely 8.5\micron). The spectral bandwidths were chosen to resemble `pseudo-continuum' ranges without any spectral lines. The FOV is 400$\times$400 mas with a pixel scale 1~mas at 3.5\micron\ and 3~mas at 8.5\micron. North is oriented up and east left, while the theoretical MATISSE beam is shown as a white circle in the left panels. All images have been convolved with a Gaussian beam with FWHM similar to the image resolution, and they are normalized to the peak intensity, and scaled in a logarithmic stretch to a minimum level of $10^{-3}$ as indicated in the colorbar. This enhances emission from the faint FUor as well as from the residuals.

In OImaging the MiRA algorithm forces the brightest component (here the HBe star) in the middle of the map, while BSMEM applies a random shift in the map. Both methods produce image reconstruction residuals due to the incomplete $uv$ coverage of the observations \citep[][ provides a thorough treatise on MATISSE image reconstructions]{planq2024}. In both cases this is manifested by multiple blobs throughout the map, while the `floral pattern' of the $uv$ coverage is seen in the MiRA reconstructions. Although BSMEM can accurately predict the location and orientation of the binary's components, the individual components have rather irregular patterns reminiscent of a patchwork (at least for the HBe star) in the non-convolved images due to the `super-resolution' effect. Similarly, MiRA presents folded-like artifacts around the HBe disk at 3.5\micron, but it is overall consistent in all bands. Overall, both methods can predict the location of the binary companion and a similar average size for both sources, therefore we trust that our image reconstruction is accurate taking into account the limited {\it uv}-coverage of our data.

\section{Optical depth variability at large scales}
\begin{figure*}
    \centering
    \includegraphics[width=0.75\linewidth]{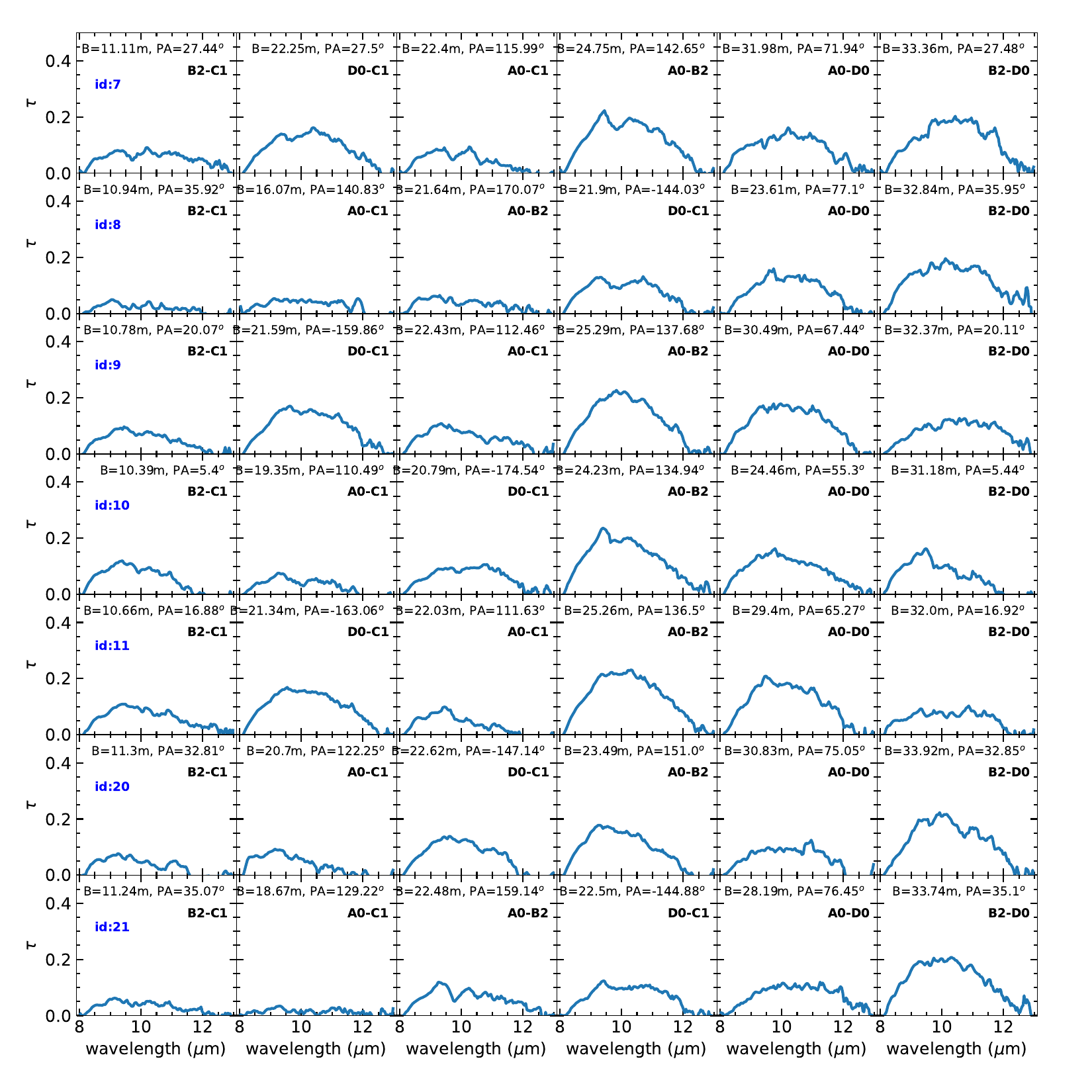}
    \caption{Optical depth ($\tau$) per baseline. The details of each baseline name, projected length and position angle, are stated within each panel. Each row corresponds to a different small-array observation and their ids (blue) can be found in Table~\ref{tab:matlog}. }
    \label{fig:alltau}
\end{figure*}

The optical depths calculated for each individual, small-array baseline as in Sect.~\ref{sec:radsil} are shown in Fig.~\ref{fig:alltau}. The spatial scales probed correspond to angular sizes of approximately 30 to 100~mas, which can be converted to physical scales of approximately 34-115~au. The binary's separation is $< 120$~mas, or else $< 135$~au and thus larger than the spatial scales probed by the small array. At the time the MATISSE observations were obtained, the pointing accuracy was not optimized to focus on any of the two stars. The imaging results show that the FUor disk is rather small and faint compared to the HBe disk, while there is no apparent extended or circumbinary warm dust emission in the MIR \citep[e.g.,][]{monnier2009}. Here, we argue that the spatial scales probed by the small array should correspond to the outer regions of the Herbig star's protoplanetary disk.

\section{Relative astrometry}\label{sec:orb}

\begin{figure*}
    \centering
    \includegraphics[width=0.8\linewidth]{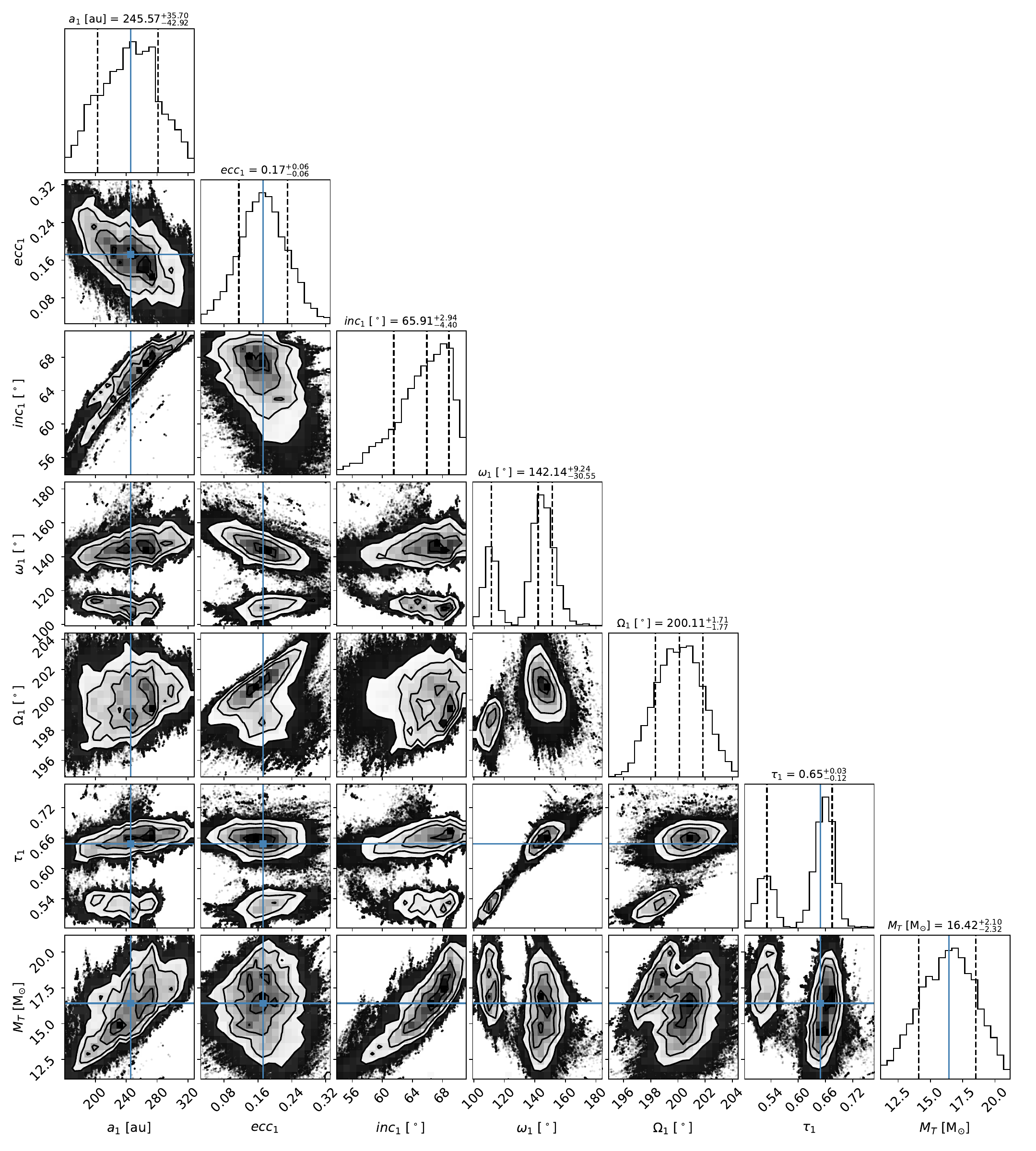}
    \caption{Corner plot of the MCMC fitting from {\tt orbitize} for the model presented in Section~\ref{sec:binary}. }
    \label{fig:corner}
\end{figure*}

Table~\ref{tab:relastro} includes the relative astrometry of the FUor star w.r.t. the HBe star from literature data, as well as our own derivation from our MATISSE observations, and the literature \citep{leinert1987, koresko1991, haas1993, malbet1993, barth1994, thiebaut1995, fischer1998, millangabet2002, ratzkaphd, szeifert2010, canovas2015, antoniucci2016, bonnefoy2017, takami2019, dong2022, thomas2023}. For the case of \citet{szeifert2010}, we provide the correct epochs of the NACO observations as extracted from the ESO archive. Following the analysis of \citet{ratzkaphd} on the 2004 MIDI/VLTI data, who modeled the binary as two uniform disk sources, we employed the same method to estimate the orbit from all archival MIDI/VLTI observations \citep{varga2018}. We supplement Table~\ref{tab:relastro} with our analysis of archival, raw, acquisition data in the NIR and MIR from NIRC2/Keck (program IDs: C216N2, H205N2L, N112), and NIRI/Gemini (program ID: GN-2010B-Q-74). For the case of NIRI/Gemini and NIRC2/Keck observations, the binary is clearly resolved even in raw acquisition data. Here, we repeat the same method as for the MATISSE images. The uncertainties are the standard deviations of the fitted separations and position angles. The analysis of the relative astrometry data is presented in Sect.~\ref{sec:binary}. Figure~\ref{fig:corner} is the MCMC fitting corner plot from {\tt orbitize}.

\begin{table}[hbtp]
    \caption{Relative astrometry of the FUor (Z~CMa SE) w.r.t. the HBe (Z~CMa NW) star.}
    \begin{tabular}{lccccc}
    \hline\hline
    Epoch (MJD) & $\rho$ (mas) & error & PA (deg) & error & Ref. \\
    \hline
46779$\dagger$ & 110.0 & 10.0 & 119.0 & 6.0 & 1 \\
47833 & 103.0 & 12.0 & 122.0 & 8.0 & 2 \\
47835 & 110.0 & 7.0 & 114.0 & 4.0 & 2 \\
48191$\dagger$ & 100.0 & 7.0 & 120.0 & 4.0 & 3,4\\
48230 & 100.0 & 10.0 & 122.0 & 2.0 & 5 \\
48264 & 100.0 & 0.1 & 120.0 & 0.1 & 6  \\
49074 & 100.0 & 8.0 & 125.0 & 2.0 & 7 \\
49382 & 100.0 & 0.01 & 122.5 & 0.2 & 4,8 \\
49792 & 105.0 & 9.0 & 124.0 & 5.0 & 9 \\
51920 & 109.0 & 2.0 & 129.0 & 0.6 & 10 \\
51920 & 108.0 & 2.0 & 129.7 & 1.3 & 10 \\
52623 & 104.5 & 2.5 & 130.89 & 1.0 & 11 \\
53340 & 109.0 & 1.0 & 130.5 & 0.3 & 12 \\
53340 & 114.4 & 2.0 & 133.1 & 2.0 & t.w. \tablefootmark{a}\\
53417 & 105.3 & 6.6 & 132.8 & 1.7 & 11 \\
53779 & 101.0 & 0.1 & 132.2 & 0.25 & 11 \\
53901 & 99.54 & 2.0 & 129.2 & 2.0 & t.w. \tablefootmark{a}\\
54821 & 109.76 & 0.53 & 131.95 & 1.43 & t.w.\tablefootmark{b}\\
54862 & 111.0 & 3.0 & 132.7 & 1.0 & 13 \\
54863 & 113.5 & 2.0 & 129.98 & 2.0 & t.w. \tablefootmark{a}\\
54888 & 111.0 & 3.0 & 132.4 & 1.0 & 13 \\
54901 & 112.0 & 3.0 & 132.1 & 1.0 & 13 \\
55172 & 111.0 & 3.0 & 133.3 & 1.0 & 13 \\
55508 & 104.5 & 2.5 & 131.85 & 0.6 & t.w.\tablefootmark{c}\\
56229 & 112.0 & 2.0 & 135.0 & 2.0 & t.w.\tablefootmark{b}\\
56936 & 110.0 & 0.1 & 136.0 & 0.1 & 14 \\
57041 & 112.0 & 3.0 & 139.6 & 2.0 & 15 \\
57112 & 114.2 & 3.1 & 136.9 & 1.5 & 16 \\
57668 & 109.0 & 4.0 & 137.4 & 2.0 & 17 \\
57682 & 116.0 & 4.0 & 140.8 & 2.0 & 17 \\
57943 & 116.0 & 4.0 & 140.8 & 2.0 & 17 \\
58003 & 116.0 & 4.0 & 140.8 & 2.0 & 17 \\
59217 & 113.18 & 1.7 & 138.17 & 0.40 & t.w.\tablefootmark{b}\\
59237* & 112.6 & 0.1 & 136.7 & 0.05 & t.w. \\
59959* & 117.88 & 0.73 & 139.16 & 0.29 & t.w. \\
    \hline
    \end{tabular}
    \tablefoot{All epochs refer to the Modified Julian Dates. ($\dagger$): Exact epoch is not known; average date derived from cited works. (*): Early GTO MATISSE data and median epoch from this work's MATISSE imaging data.
    Our analysis of archival interferometric data from \tablefoottext{a}{MIDI/VLTI}, and archival raw acquisition data from \tablefoottext{b}{NIRC2/Keck} and \tablefoottext{c}{NIRI/Gemini}.
    }
    \tablebib{(t.w.) this work; (1)~\citet{koresko1989}; (2)~\citet{thiebaut1995}; (3)~\citet{koresko1991}; (4)~\citet{thomas2023}; (5)~\citet{haas1993}; (6)~\citet{malbet1993}; (7)~\citet{barth1994}; (8)~\citet{leinert1987}; (9)~\citet{fischer1998}; (10)~\citet{millangabet2002} ; (11)~\citet{szeifert2010}, now with correct epochs as in the ESO archive ; (12)~\citet{ratzkaphd} ; (13)~\citet{bonnefoy2017}; (14)~\citet{takami2019} ; (15)~\citet{canovas2015} ; (16)~\citet{antoniucci2016}; (17)~\citet{dong2022}
    }
    \label{tab:relastro}
\end{table}

\end{appendix}
%
%
\end{document}